\documentclass[useAMS,usenatbib,usedcolumn]{mn2e}
\def\HI{\ifmmode{\rm HI}\else{H\/{\sc i}}\fi}

\def\lsun{\ifmmode{{\mathrm L}_{\odot}}\else{L$_{\odot}$}\fi}

\def\msun{\ifmmode{{\mathrm M}_{\odot}}\else{M$_{\odot}$}\fi} 
\def\msunpc2{\ifmmode{{\mathrm M}_{\odot} \, {\mathrm{pc}}^{-2}}\else{M$_{\odot} \, {\mathrm {pc}}^{-2}$}\fi}

\def\kms{\ifmmode{{\mathrm{km \, s^{-1}}}}\else{${\mathrm{km \, s^{-1}}}$}\fi}

\def\aj{AJ}%% Astronomical Journal
\def\apj{ApJ}%% Astrophysical Journal
\def\apjl{ApJ}%% Astrophysical Journal, Letters
\def\apjs{ApJS}%% Astrophysical Journal, Supplement
\def\aap{A\&A}%% Astronomy and Astrophysics
\def\mnras{MNRAS}%% Monthly Notices of the RAS
\def\pasa{PASA}%% Publications of the Astron. Soc. of Australia
\def\pasp{PASP}%% Publications of the ASP
\def\pasj{PASJ}%% Publications of the ASJ
\usepackage[figuresright]{rotating}
\usepackage{lscape}
\usepackage{graphics}
\usepackage{epsfig}
\usepackage{multirow}
\usepackage{bigdelim}
\usepackage{bigstrut}
\usepackage{amsmath}
\usepackage{amssymb}
\usepackage{lscape}

\defcitealias{Jaffe09}{J10}

\title[The effect of the environment on the gas and the stars of distant galaxies]{The effect of the environment on the gas kinematics and the structure of distant galaxies}
\author[Y.~Jaff\'e et al.] {Yara L. Jaff\'e$^{1,2,3}$\thanks{E-mail: yara.jaffe@oapd.inaf.it}, Alfonso Arag\'on-Salamanca$^1$, Harald Kuntschner$^2$, \and Steven Bamford$^1$, Carlos Hoyos$^1$, Gabriella De Lucia$^4$, Claire Halliday$^5$, \and Bo Milvang-Jensen$^6$, Bianca Poggianti$^3$, Gregory Rudnick$^7$, Roberto P. Saglia$^{8,9}$, \and Patricia Sanchez-Blazquez$^{10,11}$ and Dennis Zaritsky$^{12}$\\ 
   $^1$School of Physics and Astronomy, The University of Nottingham, University Park, Nottingham NG7 2RD, UK \\
   $^2$European Southern Observatory, Karl-Schwarzchild Strasse 2, 85748 Garching, Germany\\
   $^3$Osservatorio Astronomico di Padova INAF, vicolo dell'Osservatorio 5, I-35122 Padova, Italy\\
   $^4$Osservatorio Astronomico di Trieste INAF, Via Tiepolo 11, 34143 Trieste, Italy\\
   $^5$Osservatorio Astrofisico di Arcetri, Largo Enrico Fermi 5, I-50125 Firenze, Italy\\
   $^6$ Dark Cosmology Centre, Niels Bohr Institute, University of Copenhagen, Juliane Maries Vej 30, 2100 Copenhagen, Denmark\\
   $^7$The University of Kansas, Malott room 1082, 1251 Wescoe Hall Drive, Lawrence, KS 66045, USA\\
   $^8$Max-Planck Institut f\"{u}r extraterrestrische Physik, Giessenbachstraße, D-85741 Garching, Germany\\
   $^9$Universit\"{a}ts-Sternwarte M\"{u}nchen, Scheinerstr. 1, D-81679 M\"{u}nchen, Germany\\
   $^{10}$Departamento de Fsica Terica, Universidad Autonoma de Madrid, 28049 Madrid, Spain\\
   $^{11}$Departamento de Astrof´ısica, Universidad de La Laguna, E-38205 La Laguna, Tenerife, Spain\\
   $^{12}$Steward Observatory, University of Arizona, 933 North Cherry Avenue, Tucson, AZ 85721, USA\\}

\begin{document}

%%\date{v1.2; 26-3-2008}

\maketitle

\begin{abstract}

With the aim of distinguishing between possible physical mechanisms acting on galaxies when they fall into clusters, we study the properties of the gas and the stars in a sample of 422 emission-line galaxies from the ESO Distant Cluster Survey in different environments up to $z \sim 1$. We identify galaxies with kinematical disturbances (from emission-lines in their 2D spectra) and find that they are more frequent in clusters than in the field. The fraction of kinematically-disturbed galaxies increases with cluster velocity dispersion and decreases with distance from the cluster centre, but remains constant with projected galaxy density. We also studied morphological disturbances in the stellar light from HST/F814W images, finding that the fraction of morphologically disturbed galaxies is similar in clusters, groups, and the field. Moreover, there is little correlation between the presence of kinematically-disturbed gas and morphological distortions. For the kinematically-undisturbed galaxies, we find that the cluster and field Tully-Fisher relations are remarkably similar. In addition, we find that the kinematically-disturbed galaxies show a suppressed specific star formation rate. There is also evidence indicating that the gas disks in cluster galaxies have been truncated, and therefore their star formation is more concentrated than in low-density environments. If spirals are the progenitors of cluster S0s, our findings imply that the physical mechanism transforming cluster galaxies efficiently disturbs the star-forming gas and reduces their specific star formation rate. Moreover, this star-forming gas is either removed more efficiently from the outskirts of the galaxies or is driven towards the centre (or both). In any case, this makes any remaining star formation more centrally concentrated, helping to build the bulges of S0s. These results, in addition to the finding that the transformation mechanism does not seem to induce strong morphological disturbances on the galaxies, suggest that the physical processes involved are related to the intracluster medium, with galaxy-galaxy interactions playing only a limited role in clusters.

\end{abstract}

\begin{keywords}
Galaxies: evolution  -galaxies: clusters  -galaxies: kinematics and
dynamics -galaxies: structure
\end{keywords}

%%%%%%%%%%%%%%%%%%%%%%%%%%%%%%%%%%%%%%%%%%%%%%%%%%%%%%%%%%%%%%%%%%%%%%%%%%%%%%%
%                                                                             %
%  1. Introduction                                                            %
%  \label{sec:introduction}                                                   %
%                                                                             %
%%%%%%%%%%%%%%%%%%%%%%%%%%%%%%%%%%%%%%%%%%%%%%%%%%%%%%%%%%%%%%%%%%%%%%%%%%%%%%%
\section{Introduction}
\label{sec:introduction}

It has been well established that the fraction of spiral galaxies in clusters rises from the local universe to $z \sim 0.5$, while the S0 fraction decreases comparatively \citep{Couch1994,Dressler1997,vanDokkum1998,Fasano2000,Desai2007}. In contrast, the elliptical fraction appears to remain constant. These results imply that spirals could be transforming into S0s with time.
Moreover, galaxy morphology appears to be tightly correlated with environment and stellar populations: dense environments such as cluster cores predominantly contain galaxies with elliptical or S0 morphology \citep[$\sim80$ per cent;][]{Dressler1980,PostmanGeller1984} and very few star-forming galaxies, while the field contains a smaller fraction of galaxies that are not star-forming and of early-type morphology. In addition, the structure formation scenario of $\varLambda$CDM predicts that many galaxies have undergone the transition from field to cluster environments since $z\lesssim 1$ \citep{DeLucia2004}. 

%.

All of these results are consistent with the transformation of star-forming spirals into passive S0s by the influence of the cluster environment.  However, observational evidence has shown that galaxy clusters are not the only places where such transformation can happen. 
It is possible that  ``nurture'' is not the only driver of galaxy evolution, but that ``nature'' also plays a role: \citet{Bundy2006} suggested that there is a threshold stellar mass above which star formation is somehow quenched. These results imply that galaxy evolution might depend, at some level, on the intrinsic properties of galaxies. 
Although mass might play an important role, the stellar mass function of galaxies has been found to depend on environment \citep[][]{Baldry2006,Bolzonella2010}. Evidently, mass and environment are linked, and it is thus important to study them with caution.
It is possible that there are various physical processes responsible for the transformation of galaxies, or that different mechanisms act in different environments, but this is still unclear. 
A number of plausible mechanisms have been proposed. We summarize the most important ones here: 

\textbf{(i)} Ram-pressure stripping \citep*[][]{GunnGott1972}: the pressure due to the passage of the galaxy through the intra-cluster medium removes the galaxy's gas on timescales comparable to their cluster crossing time (a few $10^9$yr).
The HI can be removed and/or its distribution become very asymmetric, while cold molecular gas is of high enough surface density to prevent its disturbance even in the most massive clusters \citep[][for a review]{BoselliGavazzi06}. 
Depending upon the model one assumes, the gas could be removed from the disk, the halo or both, having different implications on the star formation \citep[see e.g.][]{amb99,qmb00,bcs02,BekkiCouch2003,TonnesenBryan2009,Kapferer2009,Roediger2005}

\textbf{(ii)} Mergers: simulations predict that a merger between unequal mass  spirals can form an S0 galaxy \citep{Bekki1998}, while major mergers are very likely to produce giant ellipticals \citep*[][]{NaabBurkert2003}.
In cluster cores, the high relative speeds of galaxies prevent the formation of gravitationally bound pairs during close encounters. In cluster outskirts, the environment however is less dense in general and mergers are likely to take place \citep{Mihos2003}.

\textbf{(iii)} Galaxy harassment \citep{Moore1999}: tidal forces caused by close high-speed encounters with other, more massive, galaxies can cause disk thickening and gas fueling of the central region (possibly resulting in star formation). As a consequence, the gas becomes exhausted and star formation is quenched. This mechanism is understood to be particularly important in dwarf galaxies and is most efficient in the cluster periphery.

\textbf{(iv)} Tidal interactions between galaxies and the cluster potential, ``strangulation'', or ``starvation'' \citep{Larson1980,Balogh2000}: the hot halo of a galaxy is stripped upon falling into a more massive halo. The tidal field of the cluster or group then removes the halo gas from the galaxy, halting its accretion onto the disk \citep{Bekki2001}. Hence, this mechanism effectively truncates the galaxy star formation. Although this mechanism is effective in low mass groups \citep*{McCarthy2008, KawataMulchaey2008}, it is unclear whether it can account for the apparently strong effect of the cluster environment. It is possible however, that the extreme properties observed in galaxy clusters may be the result of some ``pre-processing'' of galaxies in groups before accretion into the cluster \citep[e.g.][]{ZabludoffMulchaey1998, McGee2009}.

Each one of these mechanisms is expected to be effective in different overlapping regions of clusters, hence it can be difficult to distinguish the effects of the various physical processes with observations. Figure 1 in \citet{Moran2007} illustrates how tidal stripping is more effective towards the centre of clusters, while ram pressure stripping, starvation, and  harassment are effective out to increasingly larger radii  (in that order), and mergers dominate  outside the cluster centre. 
However, little is still  known about the importance of each mechanism to the transformation of spirals into S0s. In particular, it remains unclear whether all S0s formed through only one of the mechanisms mentioned above.

A potential difference between the various mechanisms is their predictions on the star formation within the affected galaxies. In some ram-pressure stripping models \citep[e.g.][]{BekkiCouch2003} it is possible that the star formation is enhanced across the disk, while in a merger or tidal stripping scenario,  a centrally concentrated starburst is likely to occur \citep*{MihosHernquist1994}.
However, before we can distinguish these differences, we must establish that a starburst or star formation suppression is present.

A common approach to studying the physical mechanisms driving galaxy evolution is to observe and compare the properties of well-defined galaxy samples in different environments. Examples of these properties include gas and dust content, star formation rate, chemical  composition, stellar populations, kinematics, luminosity, colour and many others.
The combination of these observables (and the ability to reproduce them with models) is crucial for a complete understanding. 
In addition to the study of individual galaxy characteristics,  understanding the effect of environment on scaling relations is a very useful way of addressing the problem. In particular, the relation between disk luminosity and maximum rotational velocity, i.e. the Tully-Fisher relation \citep*[TFR,][]{TullyFisher1977} has proven to be one of the fundamental empirical clues to the physics of galaxy formation, in particular, to the relation between dark and luminous matter in galaxies. By comparing the TFR of cluster versus (vs.) field galaxies it is possible to spot potential environmental effects that ultimately transform spirals into S0s.
Whilst the internal kinematics of galaxies reflect the overall gravitational potential (providing a proxy for the total mass), the luminosity can be used as a proxy for both luminous mass and star formation, if the right photometric band is chosen (the rest-frame $B$-band luminosity is particularly sensitive to star formation). 

Much effort has been made in understanding the local TFR, and its redshift evolution (e.g. \citealt*{TullyFisher1977}; \citealt{Cole1994,Vogt96,Ziegler2002,Kannappan2002,MJ2003,Bohm04,Bamford06,Nakamura2006,Weiner2006,Pizagno2007,Kutdemir2010}, and references therein). \citet{Kassin2007} developed a revised TFR with the aim of understanding the scatter about the stellar-mass TFR. This new relation replaced rotation velocity ($V_{\rm rot}$) with a revised kinematic estimator ($S_{0.5}$) that accounts for disordered or non-circular motions through the gas velocity dispersion $\sigma_{gal}$: $S_{0.5}^2 = 0.5V_{\rm rot}^2 + \sigma_{gal}^2$. This new relation between stellar mass and S0.5 is remarkably tight for their Keck/DEIMOS spectroscopic sample over $0.1 < z < 1.2$ with no detectable evolution in slope or intercept with redshift. They conclude from this that the galaxies are perhaps virialized over this 8 billion year period. Furthermore, they find that the S0.5 stellar-mass TFR is consistent with the absorption-line-based stellar-mass Faber-Jackson relation for nearby elliptical galaxies in terms of slope and intercept, suggesting a physical connection between them. This has also been seen locally (over a larger mass and morphology range) by \citet{Zaritsky2008}.

A few studies of the effect of the environment on the TFR have also been made. For instance, \citet{MJ2003} found, in a rather small sample (containing 8 cluster spirals at $z=0.83$ and additional field galaxies), that cluster spirals were brighter than the field ones by $\sim0.5-1$ mag at a fixed rotation velocity (1.5-2$\sigma$ result). \citet{Bamford05} found the same behavior with a significantly larger sample (111 galaxies in total at $0<z<1$). They conclude that this effect could be caused by an initial interaction with the intra-cluster medium. Controversially, \citet{ziegler03} and \citet{Nakamura2006}  found no difference between the cluster and field TFR of galaxies. Moreover, \citet{ziegler03,jager04} and \citet{metevier06} have found that in the cluster environment, many galaxies have non-circular motions. Undoubtedly, larger and more homogeneous studies that search for relations with respect to cluster properties, redshift, etc. are still needed. 

In this paper, we  use the ESO Distant Cluster Survey (EDisCS) dataset to make a statistically significant investigation of the environmental effects on galaxy evolution, by means of studying the gas kinematics, morphological disturbances, TFR, star formation, and location of the star formation within the disks of distant cluster, group, and field galaxies. The dataset is larger than all previous similar studies at high redshift and not only has the advantage of spanning a broad range in cluster properties but also contains a significant field sample to match the cluster galaxies. Unfortunately, because of the relatively low spectral resolution of our data we are not able to make a comparative study of the  S0.5 stellar-mass TFR of \citet{Kassin2007} (see Section~\ref{sec:data} for details on our dataset). Our aim is to understand which physical processes are primarily responsible for the transformation of spiral galaxies into S0s in clusters. In particular, we are interested in addressing the following questions. How is the star formation of a galaxy falling onto a cluster affected? Does it decline immediately, or does it go through a period of enhancement? If so, is there a significant offset between the cluster and field TFR? Is this last episode of star formation centrally concentrated, leading to an enhanced bulge-to-disk that would occur during a spiral-to-S0 transformation? Do these processes depend on the galaxy location within the cluster, or on cluster properties such as their mass or concentration?

The paper is organized as follows. Our dataset, galaxy selection criteria, and derived properties are described in Section~\ref{sec:data}. In Section \ref{sec:Vrot}, we explain the rotation-curve fitting procedure used to obtain reliable rotation velocities and distinguish galaxies with kinematical disturbances from the rest. We also describe the quality of the fits and our derivation of velocity measurements for each galaxy. In Section~\ref{sec:matched}, we produce  matched samples (in redshift and rest-frame B-band magnitude) that enable us to make a fair comparisons between cluster and field galaxies. We present our results in Section~\ref{sec:results}. We first quantify the fraction of galaxies with disturbed gas kinematics in different environments in Sections~\ref{subsec:badfractions} and \ref{subsec:environment}. We then perform a similar study for the morphologically disturbed galaxies in Section~\ref{subsec:morph_badfractions}. In Section~\ref{subsec:TF}, we present the TFR of cluster vs. field galaxies. 
In Section~\ref{subsec:groups}, we separate galaxy groups from clusters to study the effects that lower density environments can have on the TFR. We also explore the effect of environment on the TFR of morphologically selected spiral galaxies in Section~\ref{subsec:S_TFR}. In Section~\ref{subsec:ssfr}, we compare the specific star formation rates of Tully-Fisher galaxies with those galaxies with disturbed gas kinematics. In Section~\ref{subsec:sizes}, we examine how the location and extent of the star formation within the stellar disk is affected by environment.  We finally discuss our results in Section~\ref{sec:discusion} and draw our conclusions in Section~\ref{sec:conclusions}.

Throughout this paper, we assume a ``concordance'' $\Lambda$CDM cosmology with $\Omega_{\rm M}=0.3$, $\Omega_{\Lambda}=0.7$, and $H_{0}=70$~km~s$^{-1}$~Mpc$^{-1}$.% as in \citet{Rudnick2009}.

%%%%%%%%%%%%%%%%%%%%%%%%%%%%%%%%%%%%%%%%%%%%%%%%%%%%%%%%%%%%%%%%%%%%%%
\section{The Sample and Data}
\label{sec:data}

EDisCS is a multi-wavelength survey designed to study cluster structure and cluster galaxy evolution over a large fraction of cosmic time. The complete dataset is focused on 20 fields containing galaxy clusters at redshifts between 0.4 and 1. The cluster sample was selected to be among 30 of the highest surface brightness candidates in the Las Campanas Distant Cluster Survey \citep{Gonzalez2001}, after confirming the presence of an apparent cluster and a possible red sequence with VLT 20-min exposures in two filters. 

For the 20 fields with confirmed cluster candidates, matched optical photometry was taken using FORS2 at the Very Large Telescope (VLT) \citep[see][for a detailed description]{White2005}. The optical photometry consists of B, V, and I imaging for the 10 intermediate redshift cluster candidates and V, R, and I imaging for the 10 high redshift cluster candidates. 
In addition, near-IR J and K photometry was obtained for most clusters using SOFI at the New Technology Telescope (NTT) (Arag\'on-Salamanca et al., in preparation). 
Deep multi-slit spectroscopy with FORS2/VLT \citep{Halliday2004,MJ2008} showed that several of the confirmed fields contained multiple clusters at different redshifts \citep[cf. also][]{Gonzales2002,White2005}. 
The analysis presented in this paper is mostly based on these spectroscopic data, which we describe in some detail below.

The spectroscopic targets were selected from $I$-band catalogues \citep{Halliday2004}. Conservative rejection criteria based on photometric redshifts \citep{Pello2009} were used in the selection of spectroscopic targets to reject a significant fraction of non-members, while retaining a spectroscopic sample of cluster galaxies equivalent to a purely $I$-band selected one. \citet{Halliday2004} and \citet{MJ2008} verified that these criteria excluded at most 1.3\% of cluster galaxies. 

The extensive spectroscopic observations consist of high signal-to-noise data for $\sim 30 - 50$ members per cluster and a comparable number of field galaxies in each field down to $I \sim 22$. The wavelength ranged typically from 5300 \AA{} to 8000 \AA{} for two of the runs and 5120 \AA{} to 8450 \AA{} for the other two, although the exact wavelength range for each galaxy depends on its exact position on the mask. The exposure times were typically 4 hours for the high-z sample and 1 or 2 hours for the mid-z one. Given the long exposure times, the success rate for the spectroscopic redshifts is 97\% above the magnitude limit. The completeness of the spectroscopic catalogues, which depends on galaxy magnitude and distance from the cluster centre, was computed for each cluster in \citet{Poggianti2006}. Typically, the spectroscopy samples a region out to a cluster-centric radius equal to $R_{200}$\footnote{Where $R_{200}$ is defined as the projected radius delimiting a sphere with interior mean density 200 times the critical density, commonly used as an equivalent of virial radius.}  \citep[see][and references therein]{Poggianti2009}.

The slit size used for the spectroscopic observations was $10\times1$ arcseconds, and the spectra have a dispersion of 1.32 \AA{} pix$^{-1}$ or 1.66 \AA{} pix$^{-1}$, depending on the observing run. The masks were designed using the $I$-band images, since they best correspond to the wavelength range chosen for the spectroscopy. The slits were aligned with the major axis of the targeted object if the tilting of the slit did not exceed $\pm45^{\circ}$. In the second run however, this was only done for objects identified as late-types by the photometric redshift code (we refer to \citet{Halliday2004,MJ2008} for full details on the mask design).

The FWHM resolution of the spectroscopy is $\sim 6$ \AA{}, corresponding to rest-frame 3.8\AA{} at $z=0.6$. This translates into a rest-frame $1 \sigma$ velocity resolution of $\sim 70$ km/s at 6780 \AA{} (central wavelength of grism 600RI+19). For the typical signal-to-noise ratio in the emission lines, this means that reliable rotation velocities can be measured down to $\sim 20$ km/s.

In addition to this, ten of the highest redshift clusters from the database were enriched with Hubble Space Telescope (HST) mosaic imaging in the F814W filter with the Advanced Camera for Surveys Wide Field Camera \citep[see][for details]{Desai2007}. This allowed us to perform a visual morphological classification of the galaxies in these fields.
Moreover, three of the fields have H$\alpha$ imaging \citep{Finn2005} and three have XMM data \citep{Johnson2006}.

Cluster and field galaxies were distinguished using spectroscopic redshift
information. Galaxies whose spectroscopic redshift places them within $\pm3 \sigma_{cluster}$ of the z$_{cluster}$ in rest-frame peculiar velocity were considered cluster members. Galaxies with z outside this range were flagged as field population \citep[see][]{Halliday2004, MJ2008}. EDisCS clusters have velocity dispersion in the range $400 < \sigma_v < 1100$ km/s. Galaxy groups with velocity dispersion of $160 < \sigma_v < 400$km/s are also present \citep[See][for further details]{Poggianti2009}. 
Unless stated otherwise, the group and cluster populations will be studied together. However, in Sections~\ref{subsec:badfractions},\ref{subsec:morph_badfractions} and \ref{subsec:groups} a separate analysis of cluster, group and field galaxies will be presented.

\subsection{Structural Parameters}
\label{subsec:struc_param}

\begin{figure}
\begin{center}
  \includegraphics[width=0.5\textwidth]{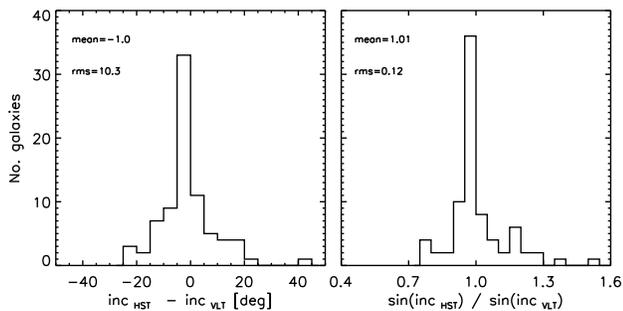}
\end{center} 
\caption{Inclinations derived from F814W HST images ($\rm inc_{\rm HST}$) are compared with those computed from $I$-band VLT images ($\rm inc_{\rm VLT}$) for galaxies within our (luminosity and redshift limited) matched samples A and B used in this work (see Section~\ref{sec:matched}). The left hand panel shows a histogram of the difference between both values. The right hand panel shows a histogram of the ratio of the sines of both inclinations. We plot these ratios to understand how much the TFR (in particular, $\log V_{\rm rot}$) would be affected. As is evident, the distribution in the right hand panel is very narrow and peaks at $1$. In both panels, the mean value and rms of the distributions are shown for reference.}
 \label{inclis}
\end{figure}

Inclinations were derived by fitting a 2-component 2D fit to F814W HST images when available, or I$_{814}$-band (VLT) images otherwise. The fit accounted for a bulge with a de Vaucouleurs profile and an exponential disk component, convolved to the PSF of the images. Disk inclinations could then be derived. This was done using the GIM2D software \citep[see][for a detailed description of the method used]{Simard2002,Simard2009}.
We verified that the use of different image data sets (HST or VLT) does not bias our results. This is illustrated in Figure~\ref{inclis}, where HST inclinations are compared with those computed from VLT images. The figure contains two histograms. The one in the left hand panel shows the distribution of the difference between the two inclinations ($\rm inc_{\rm HST}-\rm inc_{\rm VLT}$). The distribution peaks very near zero and has a rms scatter of $\sim 10$ deg. The right hand panel shows the ratio of the sines of the two inclinations, $\sin (\rm inc_{\rm HST})$ and $\sin (\rm inc_{\rm VLT})$. This was done to  quantify and understand how much the choice of one or the other value of inclination would affect the positioning of the data points on the TFR (i.e. the values of $\log V_{\rm rot}$). 
The distribution in the right hand panel is very narrow, with a clear peak at $\sin (\rm inc_{\rm HST})/\sin (\rm inc_{\rm VLT})=1$. Therefore, we can reliably use VLT-derived inclinations without biasing our results. This is also true for the less-demanding position angles.

We note that inclinations were derived from a 2D fit to the images, under the assumption that all galaxies had a ``bulge'' and a ``disk'' component \citep[see][]{Simard2009}. The presence of a ``disk'' component does not necessarily imply that there is an actual disk, because many dynamically hot systems have simple exponential profiles. In Section \ref{sec:matched}, we find that in the TFR sample (matched in $z$ and in $M_{B}$), not all the galaxies are disks, although the vast majority ($94$\%) are. 
Potential biases introduced by the small fraction of non-disk galaxies included in our sample are discussed later.

\subsection{Rest-Frame Magnitudes}
\label{subsec:rf_Mags}

The magnitudes used for the construction of the Tully-Fisher plots, and throughout this paper, were absolute B-band magnitudes, $M_{B}$. We chose $M_{B}$ because it is a good tracer of recent star formation. Values of $M_{B}$ were calculated from the observed SED of each galaxy, normalized to its total $I$-band flux, and the spectroscopic redshift \citep[we refer to][for details of the calculation of $M_{B}$ and luminosities]{Rudnick2009,Rudnick2003}.

The magnitudes were additionally corrected for internal extinction, following the prescription of \citet{Tully1998}, to give the corrected absolute rest-frame B-band magnitudes, $M_{B}$, used in this paper. 

\subsection{Star Formation Rates}
\label{subsec:SFRs}

Star formation rates (SFRs), not corrected for dust, were derived from the observed [OII]3727\AA\ fluxes following \citet{Poggianti2008}.
These fluxes were obtained by multiplying the observed [OII] equivalent width by the continuum flux, estimated from the broadband photometry using total galaxy magnitudes. 
Previous studies have shown that the obscured star forming galaxies are more common in cluster outskirts or groups \citep[e.g.][]{Saintonge2008,Gallazzi2009}. Unfortunatelly, we could not use unobscured SFRs \citep[as in ][]{Vulcani2010} due to the scarcity of mid-infrared detections in our sample.

We derived specific star formation rates (SSFRs) by simply dividing the SFRs by the stellar mass.
Stellar masses ($M_{\star}$) were computed by John Moustakas \citep[see][]{Vulcani2011} using the \textit{kcorrect} tool \citep{BlantonRoweis2007}, which models the observed broad-band photometry, fitting templates obtained with spectrophotometric models.

We used a \citet{Kroupa2001} IMF covering the $0.1\,$M$_{\odot}$--$100\,$M$_\odot$ mass range.

\subsection{The sample}
\label{subsec:TFRsample}

We focus on a sub-sample of the EDisCS dataset consisting of galaxies with measurable emission in their spectra. 
First, we rejected galaxies with emission-lines clearly affected by sky lines or without a discernible tilt (as judged by visual inspection). We then rejected galaxies with inclinations of less than $30^{\circ}$ (inclination $= 0$ corresponding to face-on) to ensure that rotation could be measured. We also rejected observations affected by slit misalignment (misalignment with respect to the major axis of the galaxy  $>30^{\circ}$) to ensure secure rotational velocity measurements. After applying these conditions, there were 1024 emission lines, belonging to a total of 422 galaxies.
Typically, we could detect 3 emission lines per galaxy. These were typically (in order of frequency), the [OII]3727\AA\ doublet, H$\beta$, the [OIII] 5007 and 4959\AA{} lines, H$\gamma$, and  H$\delta$.

The ``true'' parent emission-line galaxy distribution is well represented by our sample. 
The fraction of EDisCS galaxies with  emission-line spectra for which we were able to model emission-lines and measure a rotation curve is fairly constant ($\simeq 35$\%) in the magnitude range of our galaxies.

In Section~\ref{sec:matched}, we impose additional constraints on the sample, in both $M_{B}$ and redshift, to produce a luminosity-limited sample. This step is required before creating matched cluster and field galaxy samples. Until then, all the sample described in this section is considered, unless otherwise stated.

\begin{figure}
\begin{center}
  \includegraphics[width=0.5\textwidth]{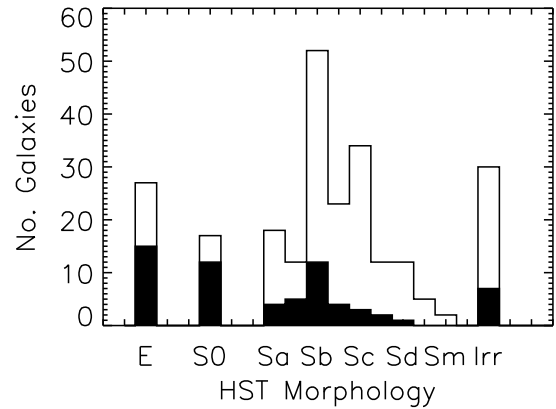}
\end{center} 
\caption{The open histogram shows the morphology distribution of the galaxies with HST observations in our measurable-emission-line sample. The shaded area will be discussed in Section~\ref{subsec:quality} and corresponds to the galaxies (within the HST sample) with poor emission line fits due to disturbed gas kinematics. The different morphologies are labeled in the plot. Whilst most of the galaxies have late-type morphologies, there is a small group of early-types in our emission-line galaxy sample.}
 \label{morph_hist}
\end{figure}

\begin{table}
\begin{center}
 \caption{Number of galaxies per morphology type for the sub-set of galaxies with HST observations. This sample is drawn from the measurable-emission-line galaxy sample, where no redshift or magnitude cuts have been made. The columns correspond to: (1) the morphology type; (2) the total number of galaxies with that morphology; and (3) the number of galaxies within that morphology group for which none of the emission-line fits were ``good'', i.e. galaxies with disturbed gas kinematics. We refer to Section~\ref{subsec:quality} for the definitions of ``good'' and ``bad'' fits. These numbers are also represented in Figures~\ref{morph_hist} and~\ref{bad_frac_morph}.} 
\label{morph_table}
\begin{tabular}{lccc}
\hline
Morphology & No. galaxies & No. ``Bad'' galaxies \\
\hline\\[-2mm]
Elliptical (E)		& 27 	& 15 	\\[1mm]
Lenticular (S0)	& 17 	& 12 	\\[1mm]
Spiral (Sa to Sm)	& 169 	& 31 	\\[1mm]
Irregular (Irr)	& 30 	& 7 	\\[1mm]
\hline 
\end{tabular} 
\\ 
\end{center}
\end{table}

As explained above, our sample selection was based on the presence of measurable emission lines (and not on galaxy morphology). It is therefore interesting to determine which galaxy morphologies passed our selection criteria. We have HST observations for 61\% of our sample, hence reliable visual morphologies \citep{Desai2007}. Figure~\ref{morph_hist} shows a histogram of the morphological types for the galaxies with HST observations. The open histogram contains all the fitted galaxies. The shaded area represents potential kinematically-disturbed galaxies, as explained later in Section \ref{subsec:quality}.

As expected, most of the emission-line galaxies in our sample are spirals, and the distribution peaks at Sb morphology types. However, somewhat unexpected, there is a significant population of early-type galaxies, 27 of which are ellipticals. We return to this finding in Section~\ref{subsec:quality} after studying the gas kinematics of the galaxies. Table~\ref{morph_table} quantifies  the morphology distribution shown in Figure~\ref{morph_hist}.

Note that in a study of the star formation histories of EDisCS galaxies, \citet{Poggianti2009} found a few spiral galaxies with spectra showing no emission lines. Obviously, these passive spirals are not present in our sample.

We also note that we are unable to identify AGNs in our data, as the traditional optical diagnostics are based on emission lines that are not included in the spectral range covered by most of our spectra. For this reason, we do not distinguish or exclude galaxies hosting an AGN from our emission-line sample. In \citet{Poggianti2008} however, it was estimated that, the contamination from pure AGNs in EDisCS spectroscopic sample is at most 7\%. Because the contamination is negligible, we  conveniently refer to galaxies interchangeably as ``emission-line'' or ``star-forming''.

%%%%%%%%%%%%%%%%%%%%%%%%%%%%%%%%%%%%%%%%%%%%%%%%%%%%%%%%%%%%%%%%%%%%%%
\section{rotational velocities}
\label{sec:Vrot}

\begin{figure*}
\centering
\hspace*{\stretch{1}}
  \includegraphics[width=1.0\textwidth]{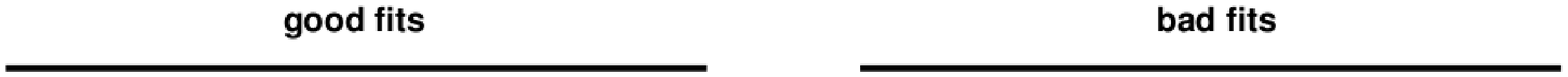}
\newline
  \includegraphics[width=0.49\textwidth]{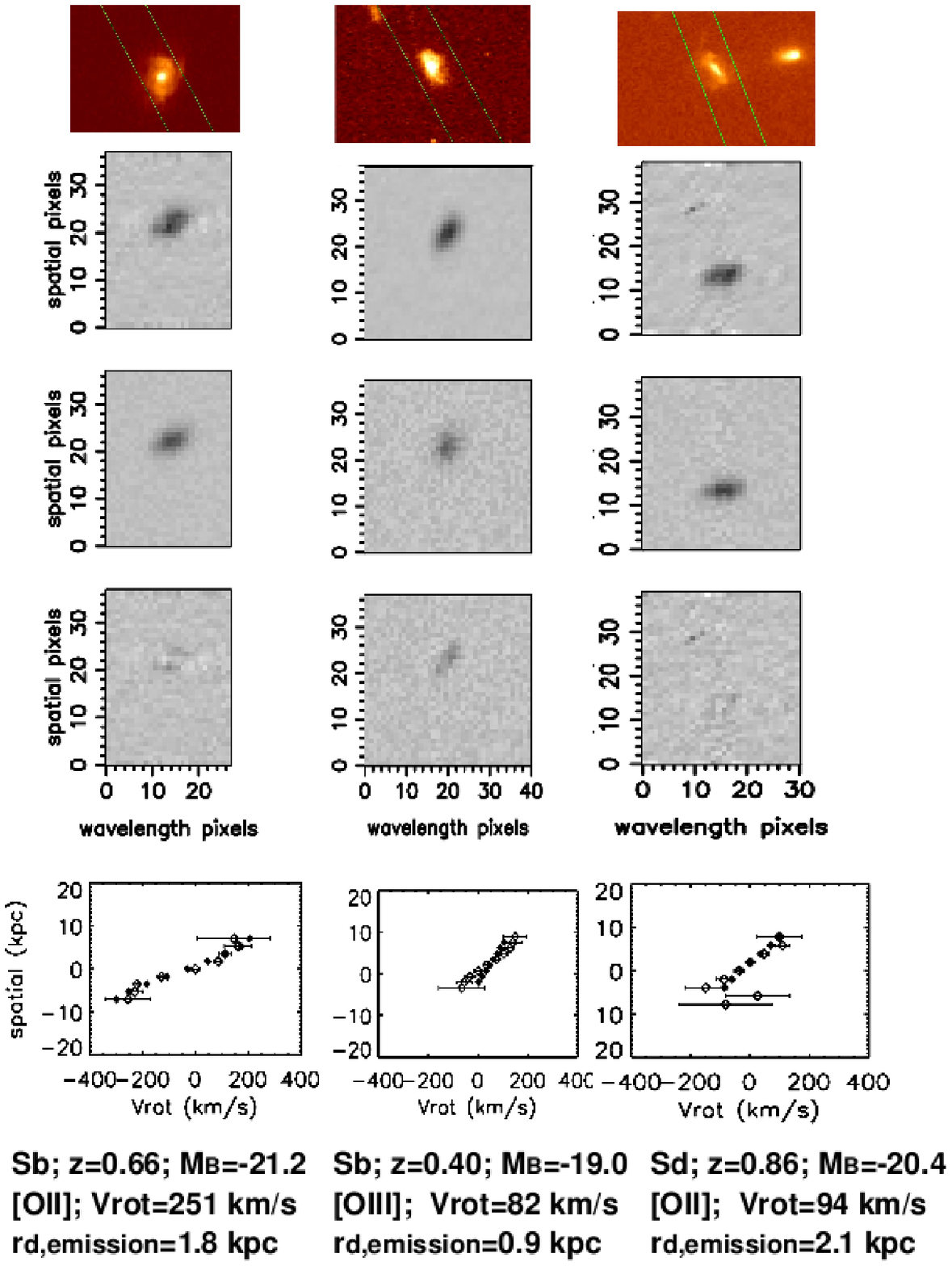}%good_f.eps}% %good1
\hspace*{\stretch{1}}
  \includegraphics[width=0.5\textwidth]{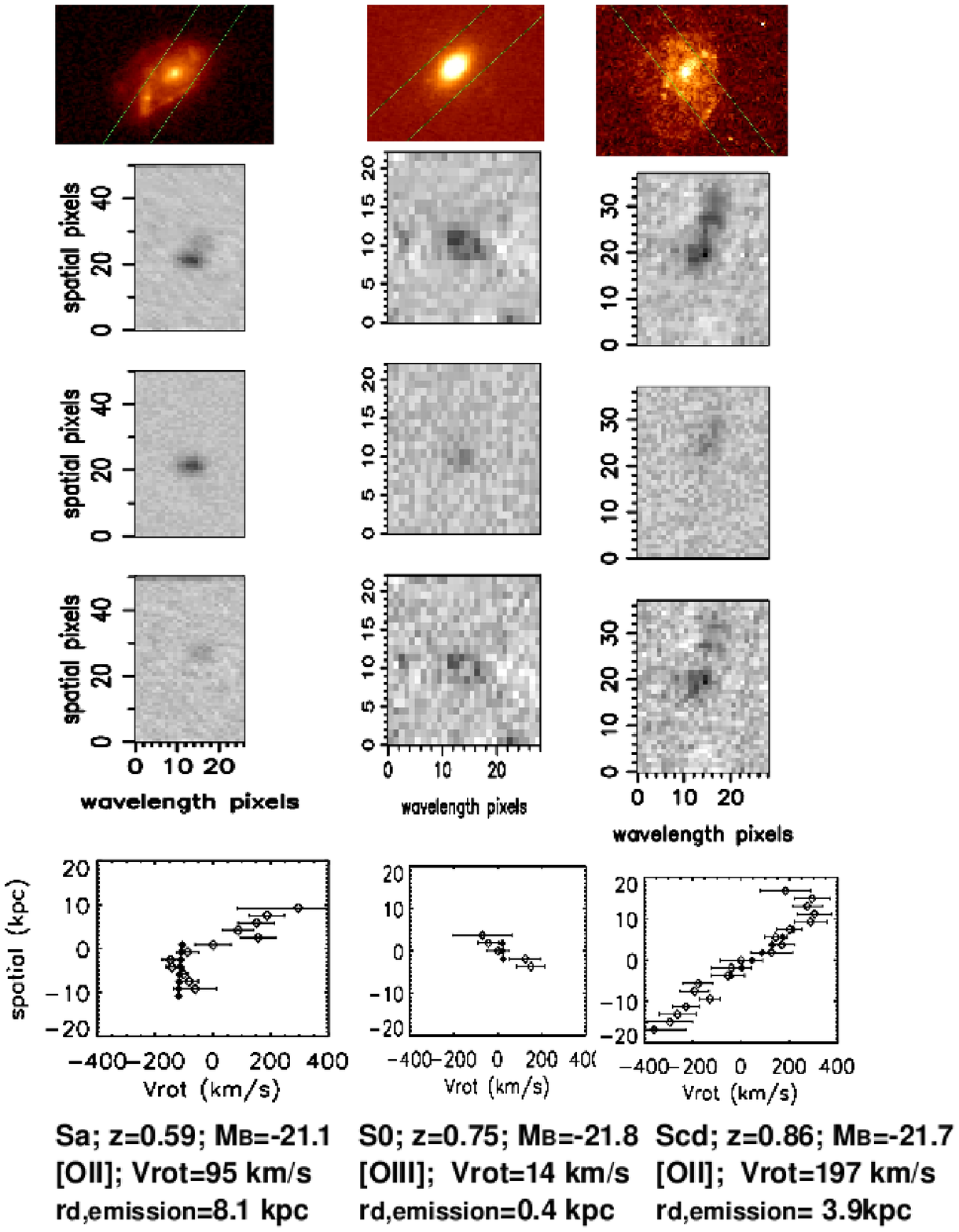}%bad_f.eps}% %good1
\hspace*{\stretch{1}}
\caption{Representative examples of our HTS images, emission line fits and rotation curves.  Our results for six galaxies are shown as means of illustrating the method and the quality of the fits. The first row shows the HST image of the galaxy with the slit position overlaid, the second row shows an extracted emission line from the 2D spectrum (postage stamp), the third row shows the best-fit model to that line, while the fourth row contains the residuals of the previous two. In addition, the traces or 1D rotation curves are shown in the bottom row in physical units.  Open circles represent the data points, while the filled ones are the model points. At the bottom of each column the morphology, redshift, $M_B$, line plotted, $V_{\rm rot}$, and $r_{\rm d,emission}$ are specified. The three panels on the left show ``good'' fits, while the three rightmost ones were classified as ``bad'' fits (see labels on the top). Note that the leftmost  panel is a very good fit, while the other two good fits (more typical) are less good but still model the data reasonably well. The bad fits on the right show clear signs of kinematical disturbance in the 2D spectra, and in the observed rotation curves. For this reason, the model fails at reproducing a \textit{Courteau} rotation curve. Also note that for example in the third column (from left to right), the emission line had a sky line subtracted. Although the subtraction is visible, this does not affect the fitting of the rotation curve significantly. There were cases however, where the sky subtraction was not as clean, making the fit a more difficult task.}
 \label{fit}
\end{figure*}

\subsection{Rotation Curve Fitting}
\label{subsec:RC}

In order to populate the Tully-Fisher diagram with trustworthy measurements, we need a reliable method to compute the rotation velocity ($V_{\rm rot}$) of the galaxies under study. We use a synthetic rotation curve method based on ELFIT2D by \citet*{SP99}, and dubbed ELFIT2PY by \citet[][]{Bamford05}, which was designed to fit rotation curves to  spatially resolved emission lines of distant galaxies. In this technique, a model emission line is created for a particular set of parameters, assuming a \textit{Courteau} rotation curve \citep[][]{Courteau1997}, and exponential surface-brightness profile. The galaxy inclination, seeing, and instrumental profile are provided as input and the fitting procedure also accounts for the galaxy size being comparable to the slit-width. A Metropolis algorithm \citep[a Markov chain Monte Carlo proposed by][]{Metropolis1953} is used to search the parameter space to find those which best fit the data, and to determine the confidence intervals of these parameters. For this work, ELFIT2PY was modified to best suit the characteristics of the EDisCS data used. Together with $V_{\rm rot,i}$, the algorithm also computes the best fit for the emission scale-length ($r_{\rm d,emission,i}$) of the line.

Because many galaxies in our sample have more than one measurable emission line, a fit was performed for each line independently. We label each line with the index $i$, which goes from 1 to the total number of emission lines available in the galaxy under study ($N$). The complete fitting procedure yielded $N$ values of $V_{\rm rot,i}$ and $r_{\rm d,emission,i}$ (as well as their uncertainties) for each galaxy. After careful quality checks (see Section~\ref{subsec:quality}), these values were then combined into unique measurements of $V_{\rm rot}$ and $r_{\rm d,emission}$ for each galaxy (see Section~\ref{subsec:unique} for details). The final errors in the measured parameters include the uncertainty caused by the multiplicity of chi-squared minima. All errors represent 68\% confidence intervals ($1 \sigma$ errors).

Final values of $r_{\rm d,emission}$ were computed using a similar procedure to that used for $V_{\rm rot}$ described in this paper. We use the emission scale-length to study the concentration of star formation with environment in Section~\ref{subsec:sizes}.

\subsection{Quality Control}
\label{subsec:quality}

To ensure the use of secure rotational velocities, we visually examined a sub-set of emission line fits and investigated whether poor fits could be identified by their reduced $\chi^2$ (output from ELFIT2PY), median and maximum signal-to-noise of the data, length of confidence intervals, and/or extent of the emission-line.  We reached the conclusion that there was no efficient way of rejecting poorly fitted emission-lines without visually inspecting their quality. For this reason, 
two people independently (YLJ and AAS) inspected the fits made to all the (1024) emission lines. We graded the fits according to their quality and created two groups: ``good'' and ``bad''. 
Both classifiers agreed in the vast majority of the cases (91\%). In the few cases where we disagreed, we adopted the most pessimistic outcome.
This classification yielded 521 ``good'' quality fits (i.e. reliable emission line fits) and 503 ``bad'' ones.
The ``bad'' fits correspond to either lines with poor signal, artefacts in the postage stamps (e.g. a poorly subtracted overlapping sky line or cosmic rays), or more frequently, poor fits due to  disturbed gas kinematics in the targeted galaxy (i.e. observed rotation curve that did not resemble a rotating disk). We note that generally, galaxies with kinematically ``bad'' fits consistently showed the same distorted features in all their visible emission lines.
The ``bad'' fits were not used in the Tully-Fisher analysis (Section~\ref{subsec:TF}). However, we used the information that they provided in a parallel study of the fraction of potential ``kinematically disturbed'' galaxies with luminosity and environment (see Sections~\ref{subsec:badfractions} and ~\ref{subsec:environment}). After rejecting the ``bad'' fits, our sample decreased in size to 521 lines belonging to 289 galaxies. By performing such sample cleaning, we are able to ensure that all the fits used have reliable rotation curves, hence reliable measurements of $V_{\rm rot}$. Figure~\ref{fit} shows examples of ``good'' and ``bad'' emission-line fits. \footnote{The complete set of emission line fits can be found in the EDisCS website at: http://www.mpa-garching.mpg.de/ediscs/Papers/Jaffe\_tfr\_2011/RCfits.html}.
%http://www.nottingham.ac.uk/$\sim$ppxyj/RCfits.html}.
More than half of the galaxies (55\%) had more than one ``good'' emission line. The remaining galaxies had only one measurable emission line from which a final rotational velocity could be computed.

\begin{figure}
\begin{center}
  \includegraphics[width=0.49\textwidth]{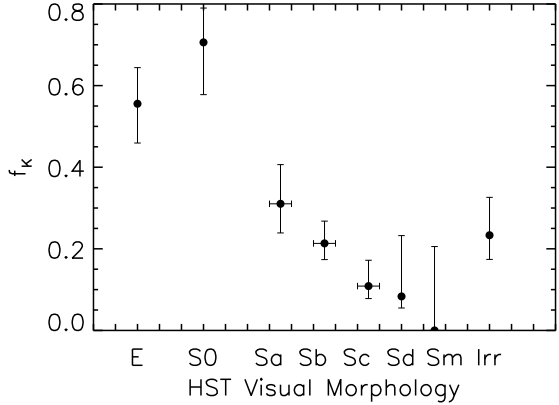}
\end{center} 
\caption{The fraction of galaxies with disturbed kinematics ($f_{\rm K}$) in the whole emission-line sample is shown for the different morphologies. Galaxies in cluster/group and the field are plotted. Horizontal error bars (when present) represent the grouping of adjacent morphology types, these cases are: Sa + Sab, Sb + Sbc, and Sc + Scd. This was done to increase the number of galaxies in these morphology bins.
Error bars are the confidence intervals (c$\approx$0.683) for binomial populations, from a beta distribution \citep[see][]{Cameron2011}.}
 \label{bad_frac_morph}
\end{figure}

In Section~\ref{subsec:TFRsample}, we showed that the parent emission-line sample spans a wide range of morphologies but is mostly composed of spirals. At this stage, it is interesting to see how the quality of the emission-line fits is correlated with morphology, especially if we assume that galaxies with ``bad'' fits are kinematically disturbed systems. The shaded area in the histogram of Figure~\ref{morph_hist} shows the morphology distribution of the poorly fitted galaxies (galaxies for which all the emission-line fits available were flagged as ``bad''). The open histogram draws the distribution of the full (good and bad) parent sample where HST images were available. 
The fraction of ``bad'' or kinematically-disturbed galaxies ($f_{\rm K}$) is plotted as a function of morphology in Figure~\ref{bad_frac_morph}, and Table~\ref{morph_table} lists (in numbers) the amount of ``bad'' fits obtained  for each morphology group. 
It is evident that the worst fitted group of galaxies (the ones showing the greatest deviations from a \textit{Courteau} rotation curve) are the early types (E and S0s). Interestingly, the worst fitted galaxies seem to be S0s and not the ellipticals nor the irregulars. However, as expected, the galaxies with the least amount of ``bad'' fits are the spirals. In the context of spiral-to-S0 transformation, this implies that galaxies already having S0 morphology have been subjected to strong disturbances in their gas content.

A very interesting finding is the discovery of 41 emission-line early-type galaxies, 17 of which have ``good'' rotation curve fits. These galaxies could be the first observational evidence of the existence of intermediate redshift early-type galaxies with a gaseous extended rotating disk. We will pursue a separate analysis on this interesting group of galaxies in a future paper (Jaff\'e et al., in preparation).

\subsection{Unique measurements of $V_{\rm rot}$}
\label{subsec:unique}

We combined the individual rotational velocity measurements in each galaxy into a unique $V_{\rm rot}$ value taking only ``good'' quality fits into account. After performing the fits and the quality checks (Sections~\ref{subsec:RC} and \ref{subsec:quality}) , we were left with $N_{\rm good}$ values of $V_{ \rm rot,i} \pm ^{\sigma_{i}^{+}}_{\sigma_{i}^{-}} $ per galaxy, where the index $i$ represents the individual lines and $\sigma_{i}^{-}$ and $\sigma_{i}^{+}$ are the left- and right-hand side errors in $V_{ \rm rot,i}$. These (asymmetric) errors come from the best-fit model's confidence intervals. We then combined the $V_{ \rm rot,i}$'s by taking the weighted average, given by
\begin{equation}
V_{ \rm rot}=\dfrac{\sum_{i=0}^{N_{\rm good}} \omega_{i} V_{ \rm rot,i}}{\omega_{i}},
\label{eq:wmV}
\end{equation}
where $\omega_{i}=1/\sigma^{2}_{tot,i}$, and  $\sigma^{2}_{tot,i}=[(\sigma_{i}^{+})^{2} + (\sigma_{i}^{-})^{2}]/2$, i.e. the average variance. 
The  upper and lower errors ($\sigma^{+}_{V_{ \rm rot}}$ and $\sigma^{-}_{V_{ \rm rot}}$, or just $\sigma^{\pm}_{V_{ \rm rot}}$) in the unique $V_{ \rm rot}$ were also evaluated by combining the individual errors in each galaxy. 
These unique error values were determined as the maximum value of the following two quantities:
\newline (i) A weighted combination of the standard errors ( $\sigma_{i}^{\pm}$) estimated by the best-fit model
\begin{equation}
\sigma_{V_{ \rm rot,com}}^{\pm} = \sum_{i=0}^{N_{\rm good}} \sigma_{i}^{\pm} \left(\frac{\omega_{i}}{\sum_{i=0}^{N_{\rm good}} \omega_{i}}\right)^{2};
\label{eq:(i)}
\end{equation}
\newline (ii) The standard error in the weighted mean, determined from the individual measurements
\begin{equation}
(s_{V_{ \rm rot}})^2=\frac{\sum_{i=0}^{N_{\rm good}} (V_{ \rm rot,i} - V_{ \rm rot})^2}{N_{\rm good}-1}.
\label{eq:(ii)}
\end{equation}
\newline In other words, the $+$ and $-$ errors in $V_{ \rm rot}$ are given by
\begin{equation}
(\sigma^{\pm}_{V_{ \rm rot}})^{2} = Max \binom{(\sigma_{V_{ \rm rot,com}}^{\pm})^{2}}{(s_{V_{ \rm rot}})^{2}}.
\label{eq:max}
\end{equation}

In this way, we take into account the cases for which there were inconsistent velocity measurements within galaxies with more than one emission line. In these cases, Equation~\ref{eq:(i)} would underestimate the true uncertainty, while Equation~\ref{eq:(ii)}   provides a more realistic error. The only problem in using the described $Max$ function (Equation~\ref{eq:max}) arises for galaxies with only one measured emission line for which Equation~\ref{eq:(ii)} has no meaning. However, we consider this to be a minor problem compared to the possibility of seriously underestimating the uncertainties. In most cases (66\% of the time), Equation~\ref{eq:max} yielded $(\sigma^{\pm}_{V_{ \rm rot}})^{2}=(\sigma_{V_{ \rm rot,com}}^{\pm})^{2}$.

\begin{figure}
\begin{center}
  \includegraphics[width=0.5\textwidth]{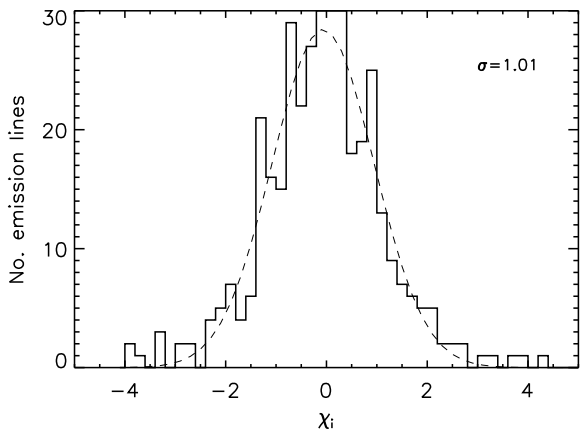}
\end{center} 
\caption{A histogram of the computed $\chi_{i}$ (see Equation~\ref{chi_eq}) for the independent velocity measurements in the galaxies with more than one good emission line available. The Gaussianity of the $\chi_{i}$-distribution and its unity standard deviation provides a high degree of confidence in the total errors in the rotational velocities used in this work.}
 \label{chi}
\end{figure}

To test the reliability of the measured errors we also computed a $\chi_{i}$ for each value of $V_{\rm rot,i}$ by calculating the quantity
\begin{equation}
\chi_{i}=\frac{V_{\rm rot,i}-V_{\rm rot}}{\sqrt{\frac{(\sigma_{\rm i}^{+})^{2} + (\sigma_{\rm i}^{-})^{2}}{2} + \frac{(\sigma^{+}_{V_{ \rm rot}})^{2} + (\sigma^{-}_{V_{ \rm rot}})^{2}}{2} }},
\label{chi_eq}
\end{equation}
which  has a physical meaning only for galaxies with more than one velocity measurement. Figure~\ref{chi} shows a histogram of the $\chi_{i}$ values obtained. As is clearly evident, the $\chi_{i}$  distribution is very Gaussian and has a standard deviation remarkably close to 1, giving a high degree of confidence in the total errors used in this work and confirming that our errors are internally consistent.

A complete table with the final $V_{\rm rot}$, $r_{\rm d,emission}$, and other characteristics of our full sample can be found in the electronic version. In this table, we have flagged the galaxies for which we had good or bad emission-line fits. We note that the galaxies that did not have a meaningful fit because they only had ``bad'' emission-line fits still have listed values of $V_{\rm rot}$ and  $r_{\rm d,emission}$ and special care should be taken in using this numbers. Table~\ref{example_table} in the Appendix shows as an example, 10 (arbitrarily chosen) lines of the complete table.

%%%%%%%%%%%%%%%%%%%%%%%%%%%%%%%%%%%%%%%%%%%%%%%%%%%%%%%%%%%%%%%%%%%%%%
\section{Matched Samples}
\label{sec:matched}

\begin{figure*}
\begin{center}
  \includegraphics[width=0.85\textwidth]{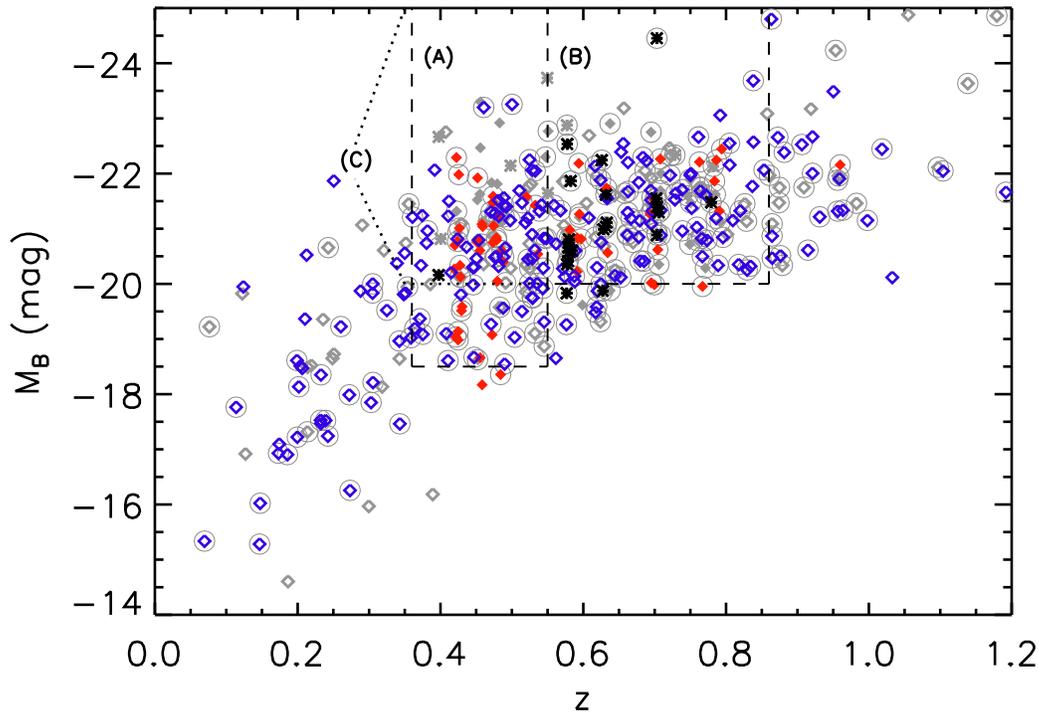}
\end{center} 
\caption{Distribution of $M_{B}$ with redshift for the 422 galaxies of our measurable-emission-line sample. The cluster galaxies are plotted in filled diamonds, groups ($\sigma_{cl}<400$km/s) are represented as asterisks and the field sample corresponds to the open diamonds. The red, black and blue colours (for cluster, group and field galaxies respectively) correspond to those galaxies with ``good'' rotation-curve fits, whilst the grey symbols represent the poorly-fitted galaxies. Three sub-samples are drawn from this plot: the lower redshift matched sample A (labeled dashed-line box), the higher redshift sample, B (again drawn within a dashed-line box), and an overall matched sample C that covers the redshift range $0.36<z<0.75$ and has the same magnitude limit as sample B (see dotted lines for guidance). For future reference, we have highlighted galaxies with HST observations with a surrounding grey circle.}
 \label{matched_sample}
\end{figure*}

Our emission-line galaxies (with fitted rotation curves) span a broad range of redshifts and rest-frame $B$-magnitudes, as Figure~\ref{matched_sample} shows. Galaxies of all qualities are plotted. The galaxies with ``good'' rotation curve fits are plotted in colours other than grey, depending on their environment:  the blue open diamonds correspond to field galaxies, red filled diamonds to cluster galaxies, and black asterisks to group galaxies. The galaxies with ``bad'' rotation-curve fits are plotted in grey with the same symbols for environment. It is clear that there are more field than cluster/group galaxies ($\sim$70\% of the emission-line galaxies are in the field).  Field galaxies are also more widely distributed in both redshift ($0<z<1.2$) and  rest-frame $B$-magnitudes ($M_{B}<-14$) than the cluster/group population. The difference in redshift between the field and cluster sample is a direct result of the redshift of our clusters. The different ranges in $M_{B}$ are a consequence of the different redshift ranges, as the observed $I$-band  targeting limit was the same for both cluster and field galaxies.

To investigate possible differences between cluster/group and field galaxies, we created field galaxy samples to match the cluster/group population. We did this by imposing cuts in redshift and $M_{B}$ simultaneously.
Three different cuts were made, producing three luminosity-limited or ``matched'' samples, represented (with boxes) in Figure~\ref{matched_sample} and summarized in Table~\ref{matched_table}.

The samples containing all (``good'' and ``bad'') galaxies are used in in Sections~\ref{subsec:badfractions}, ~\ref{subsec:environment}, ~\ref{subsec:morph_badfractions} and ~\ref{subsec:ssfr}, while in Sections~\ref{subsec:TF},~\ref{subsec:groups}, and ~\ref{subsec:S_TFR} only the matched samples containing galaxies with good rotation-curve fits, and velocities consistent with rotation, are considered.

\begin{table*}
\begin{center}
 \caption{Characteristics of the matched samples A, B and C of cluster/group and field galaxies (see Figure~\ref{matched_sample}), as well as for the whole sample (without any $M_{B}$ or $z$ cuts). For each sample, the following information is given: the magnitude limit, redshift range, number of galaxies with ``good'' rotation-curve fits, %number of galaxies with  ``good'' rotation-curve fits and velocities consistent with non-zero rotation,
 and number of galaxies with poor or ``bad'' rotation-curve fits. The last two quantities are given for cluster/group galaxies (labeled ``Cluster'') as well as for galaxies in cluster/groups or the field (labeled ``Total''). The number of field galaxies in each case can be calculated by simply subtracting the number of cluster/group galaxies from the total number. The table also gives the sample sizes for the sub-samples with HST observations, in the same format as explained above.} 
\label{matched_table}
\begin{tabular}{lcccc}
\hline\\[-2mm]
			&\textbf{ Sample A} 			& \textbf{Sample B}	& \textbf{Sample C} 		& \textbf{No cuts}\\[2mm]
\hline\\[-2mm]
$M_B$ (faint) limit	& -18.5 mag			& -20.0	mag			& -20.0 mag		& - \\[1mm]
redshift range		& $0.36 \leq z \leq 0.55$	& $0.55 < z \leq 0.86$	& $0.36 \leq z \leq 0.86$& - \\[1mm]
\hline \\ [-2mm]
\textbf{All galaxies}		& Cluster \hspace{2mm} Total		&  Cluster \hspace{2mm} Total		&  Cluster \hspace{2mm} Total		& Cluster \hspace{2mm} Total	\\[2mm]
Total No.			&\hspace{2mm}57 \hspace{5mm} 143 	&\hspace{3mm}60 \hspace{5mm} 151 	&\hspace{2mm}109 \hspace{5mm}  264	& 132 \hspace{5mm} 422 \\[2mm]
No. ``good'' galaxies		&\hspace{2mm}35 \hspace{5mm}  100	&\hspace{2mm}37 \hspace{5mm} 105	&\hspace{2mm}65 \hspace{5mm}  181	&\hspace{2mm}81 \hspace{5mm}  289 \\[1mm]
%No. ``good'' \& rotating galaxies&\hspace{2mm}30 \hspace{5mm}  92 	&\hspace{2mm}27 \hspace{5mm}  59	&\hspace{2mm}XX \hspace{5mm}  133 	&\hspace{2mm}69 \hspace{5mm}  247 \\[1mm]
No. ``bad'' galaxies		&\hspace{1mm}22 \hspace{5mm}  43	&\hspace{1mm}23 \hspace{5mm} 46 	&\hspace{1mm}44 \hspace{6mm}  83	&\hspace{2mm}51 \hspace{5mm}  133 \\[1mm]
\hline \\ [-2mm]
\textbf{Galaxies with HST observations:}        & Cluster \hspace{2mm} Total		&  Cluster \hspace{2mm} Total		&  Cluster \hspace{2mm} Total		& Cluster \hspace{2mm} Total	\\[2mm]
Total No.					&\hspace{2mm}23 \hspace{5mm} 69 	&\hspace{2mm}56 \hspace{5mm} 111 	&\hspace{2mm}73 \hspace{5mm}  155	&\hspace{1mm} 88 \hspace{5mm} 259 \\[2mm]
No. ``good'' galaxies				&\hspace{1mm} 18 \hspace{5mm} 55	&\hspace{2mm} 34 \hspace{5mm} 77	&\hspace{0mm} 47 \hspace{6mm} 112	&\hspace{1mm} 59 \hspace{5mm} 188 \\[1mm]
%No. ``good'' \& rotating galaxies		&\hspace{2mm} XX \hspace{5mm} YY 	&\hspace{2mm} XX \hspace{5mm} YY	&\hspace{2mm}  XX \hspace{5mm} YY	&\hspace{2mm} XX \hspace{5mm} YY \\[1mm]
No. ``bad'' galaxies				&\hspace{2mm} 5 \hspace{5mm} 14		&\hspace{2mm} 22 \hspace{5mm} 34	&\hspace{0mm}  26 \hspace{6mm} 43	&\hspace{0mm} 29 \hspace{5mm} 72 \\[1mm]
\hline 
\end{tabular} 
\\ 
\end{center}
\end{table*}

The redshift cuts for samples A and B were chosen so that each bin spans a similar amount of cosmic time ($\sim 1.5$ Gyr). Therefore, sample C spans $\sim 3$ Gyr of cosmic time.
In what follows, we only consider galaxies within the limits of these 3 matched samples, unless otherwise stated. By doing this, we ensure a fair comparison between field and cluster galaxies (similar epochs and luminosities), which is the main goal of this paper. 

\begin{figure}
\begin{center}
  \includegraphics[width=0.5\textwidth]{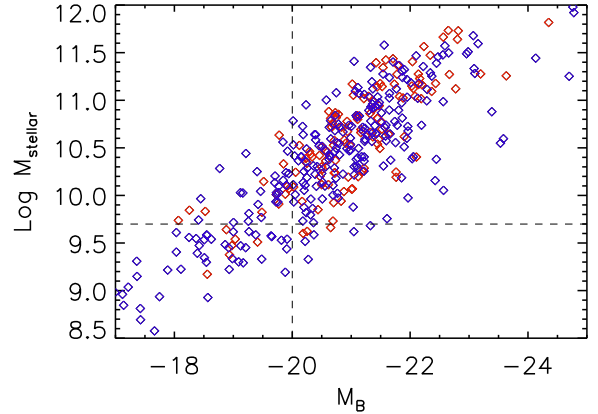}
\end{center} 
\caption{$B$-band magnitude is plotted against the logarithm of the stellar mass for those emission-line galaxies in the range $0.36 \leqslant z \leqslant 0.86$. This plot shows that because $M_B$  and $M_{\star}$ are clearly correlated, a stallar mass selection  would not difeer much from a magnitude selection. The $M_{\rm B}$ limit of sample C is shown in a vertical dashed line and a close-equivalent $M_{\star}$ limit is shown in a horizontal dashed line.}
 \label{mass_mag}
\end{figure}

We created matched samples in $M_{B}$ rather than in stellar mass ($M_{\star}$), to keep the sample selection as close to the observables as possible. We note however that matching the samples in $M_{\star}$ does not make a significant difference since $M_B$ and  $M_{\star}$ are well correlated in our sample (see Figure~\ref{mass_mag}). Our $M_{B}$-matched sample C is equivalent to a $M_{\star}$-matched sample of $M_{\star}\gtrsim3\times10^9$M$_{\odot}$, in the same redshift range, with the exception of a few galaxies ($\sim2$\% of the galaxies in C). Figure~\ref{mass_mag} shows that, although there is some scatter, $M_B$  and $M_{\star}$ are clearly correlated. In this plot, the $M_B$  limit is shown as a vertical dashed line, and the $M_{\star}$ limit as a horizontal one. These lines delimit four regions in the plot: the upper-right region contains galaxies selected in both magnitude and mass (73.7\%), the upper-left area contains those selected in mass but not in magnitude (9.7\%), the lower-right region those selected in magnitude but not in mass (2.4\%), and the lower-left those not selected in neither mass or magnitude (14.2\%).

\begin{figure}
\begin{center}
  \includegraphics[width=0.5\textwidth]{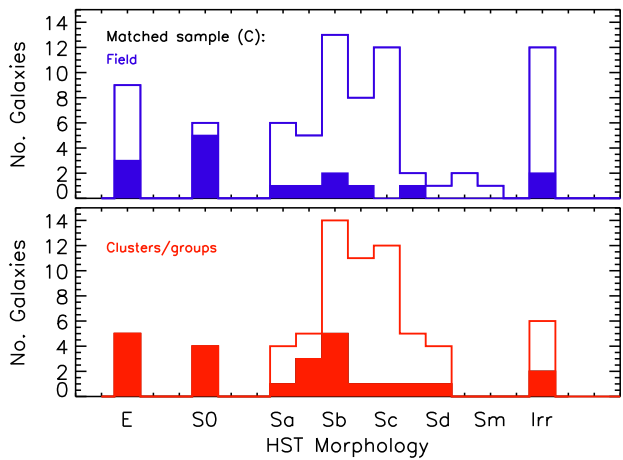}
\end{center} 
\caption{Histogram of the morphological types for the galaxies with HST observations that are in the C matched sample for field (top panel) and the cluster/groups (bottom panel). The filled areas (in both panels) represent the galaxies for which all rotation-curve fits were ``bad''.  The different morphologies are labeled in the plot.}
 \label{matched_morph_histo_clf}
\end{figure}

The morphologies of the cluster and field galaxies in the matched sample C are shown in Figure~\ref{matched_morph_histo_clf} for galaxies with HST observations. The filled areas correspond to galaxies with ``bad'' rotation curve fits or disturbed gas kinematics in field (upper panel) and cluster/group (lower panel) environments respectively. The overall distribution (of ``good'' plus ``bad'' galaxies) is shown in the solid lines (open histograms) in each case. Although the numbers are low (due to the sample being restricted to HST observations),  the figure shows that there are more ``bad'' fits in cluster environments ($\sim44\%$) than in the field ($\sim25\%$). This effect is studied thoroughly in Section~\ref{subsec:badfractions} for the matched sample C. Figure~\ref{matched_morph_histo_clf} also shows that while all of the cluster/group early-type galaxies had ``bad'' fits, 7 field  early-types (6 ellipticals and 1 S0) in this ``matched'' sample survived the quality filters. We emphasize that in the morphology distribution shown previously in Figure~\ref{morph_hist}, the number counts are higher than in Figure~\ref{matched_morph_histo_clf} because in Figure~\ref{morph_hist}, we did not restrict our emission-line sample in any way, whilst in Figure~\ref{matched_morph_histo_clf} we imposed magnitude and redshift cuts to create a ``matched'' sample. 

As mentioned in Section~\ref{sec:data} when describing the data, the rest-frame $B$-band magnitudes were corrected for internal extinction. When accounting for this effect, we used the galaxy inclinations, which were calculated from the measured ellipticities, assuming all the galaxies to be disks. As  Figure~\ref{matched_morph_histo_clf} illustrates, not all of the galaxies in our matched sample are disks. We note however that the number of ellipticals is so small (6 with ``good'' fits in samples A and B) that the $M_B$ correction applied to them does not alter our results. However, the inclination correction could potentially underestimate the luminosity and this may produce scatter in the TFR (Section~\ref{subsec:TF}) since both $M_{B}$ and the rotational velocity depend on the inclination. The typical $M_{B}$ correction for these galaxies was very small ($\sim -0.3$ mag), since their inclinations were all below $\sim 55^{\circ}$. In Section~\ref{subsec:S_TFR} however, we study the TFR of (strictly) morphologically selected spirals, where the inclination correction is more reliable.

%%%%%%%%%%%%%%%%%%%%%%%%%%%%%%%%%%%%%%%%%%%%%%%%%%%%%%%%%%%%%%%%%%%%%%

\section{Results}
\label{sec:results}

\subsection{Kinematically Disturbed Galaxies}
\label{subsec:badfractions}

\begin{figure*}
\begin{center}
  \includegraphics[width=0.49\textwidth]{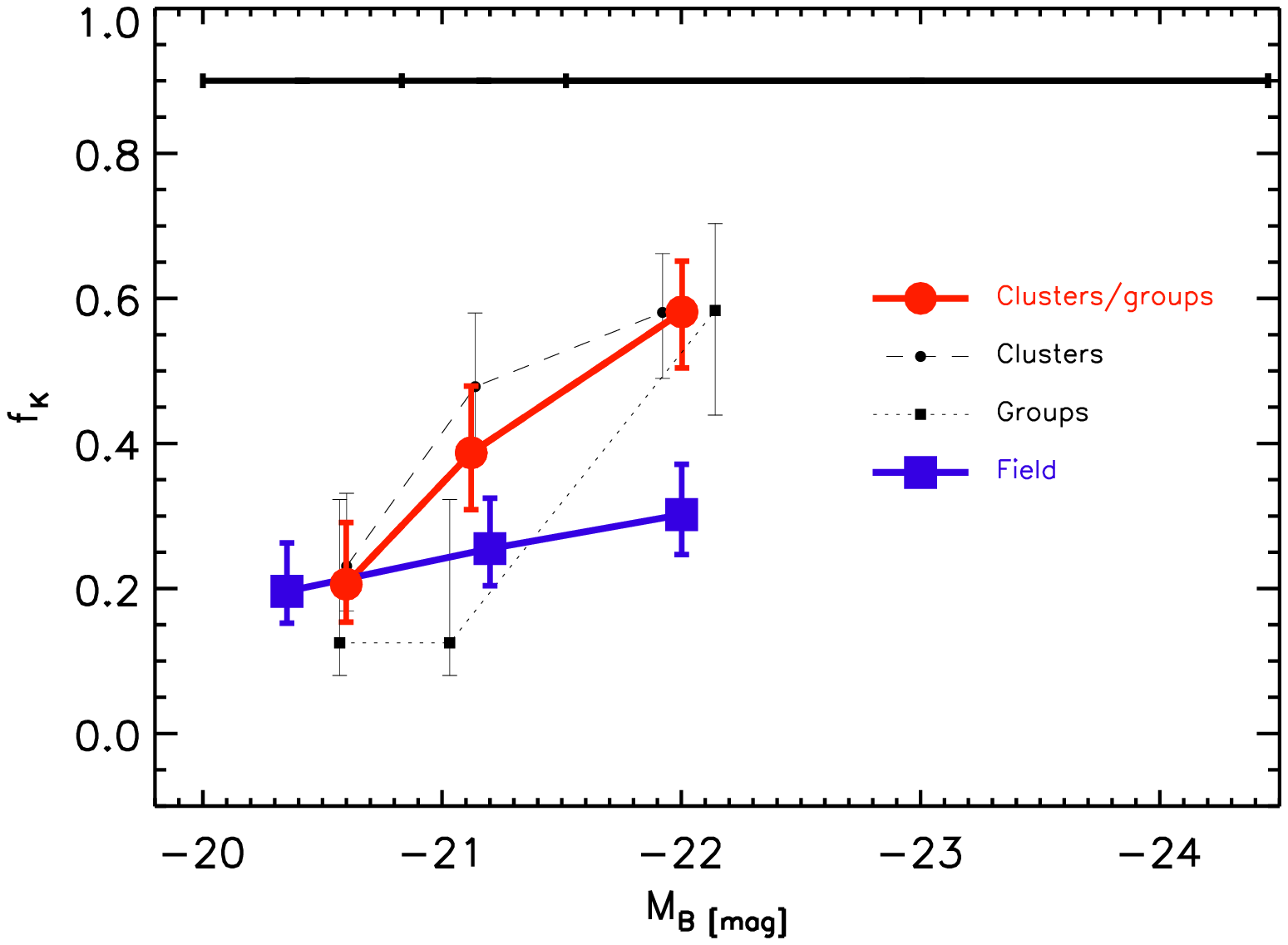}
 \includegraphics[width=0.49\textwidth]{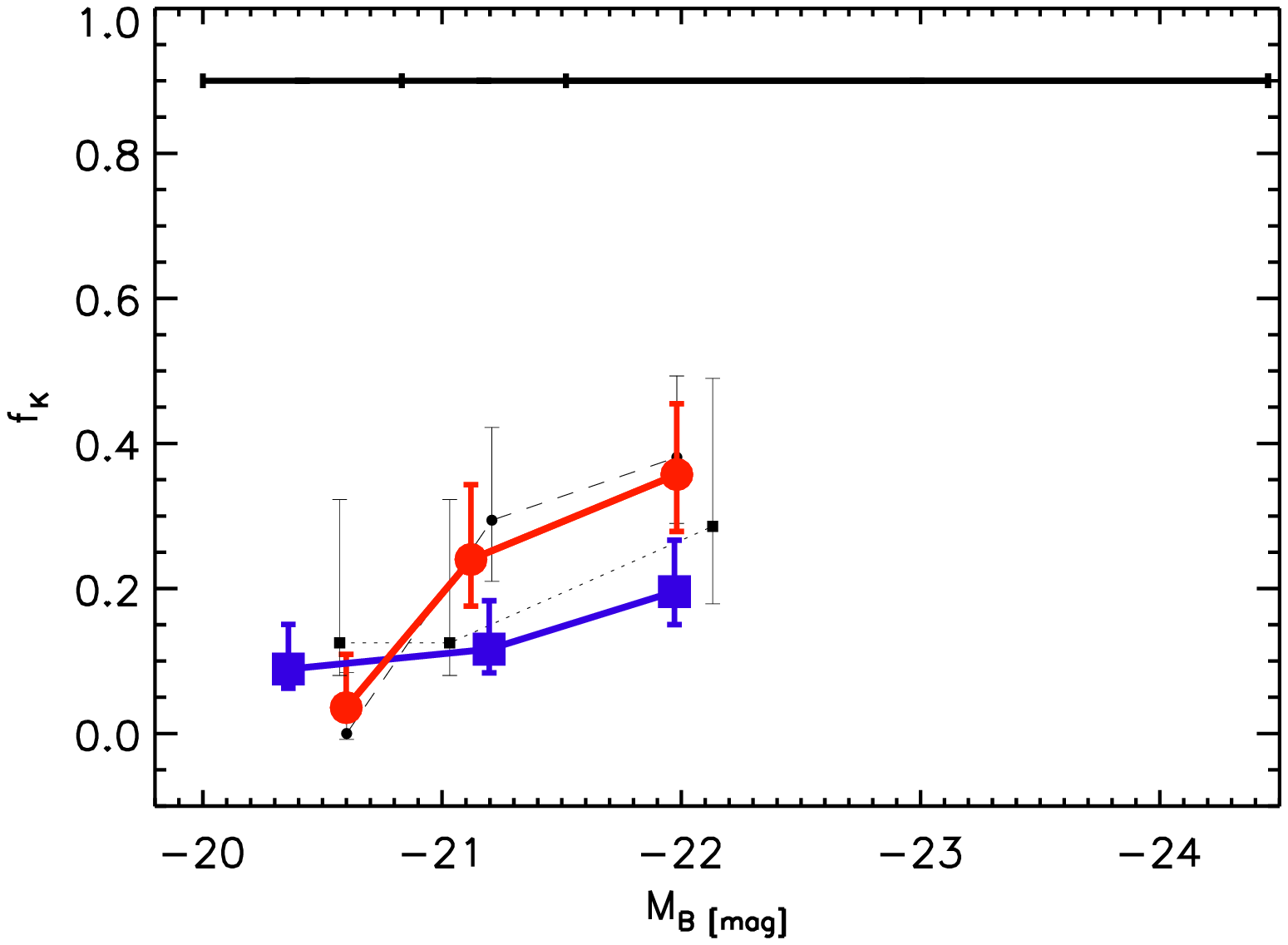}
\end{center} 
\caption{The fraction of galaxies with disturbed kinematics is shown for different environments as a function of $M_{B}$ for the ``matched'' sample C (see Figure~\ref{matched_sample} or Table~\ref{matched_table} for the definition of the samples). In the left hand panel, all the ``bad''  rotation curve fits are considered. The right hand panel shows the same but with a more conservative cut in the definition of ``bad''. In this case, we have revised the ``bad'' emission-line fits to isolate galaxies with ``secure'' kinematical distortions (see text for details) and reject galaxies with spectra that are presumably affected by artefacts. The different environments are shown in the legend of the plot on the left and also apply for the right hand plot. The error bars in the abscissa correspond to  confidence intervals for binomial populations  \citep[from a beta distribution, see][]{Cameron2011} and the horizontal error bars (shown at the top of the plots) simply represent the bin size in $M_{B}$. These bins were chosen to contain similar number of field galaxies. The position of the points correspond to the median value of the galaxies within their magnitude bin. It is clear that the plot on the right agrees with the plot on the left, albeit with larger error bars due to the reduced sample size.}
 \label{badfrac}
\end{figure*}

As explained in Section~\ref{subsec:quality}, a significant fraction of the fits made to the emission lines in our galaxy sample were classified as ``bad'' fits. Many of these lines showed evidence of disturbed gas kinematics in the galaxy, thus, a \textit{Courteau} rotation curve could not provide a good fit. We use this information to investigate the fraction of galaxies with disturbed gas kinematics (``bad'' galaxies) with environment. The left-hand panel of Figure~\ref{badfrac} shows the fraction of ``bad'' over total number of galaxies ($f_{\rm K}=\frac{N_{\rm bad}}{N_{tot}}$) in the matched sample C as a function of $M_{B}$ (in bins that contain the same number of field galaxies). 
Although sample C spans a broad redshift range, in Section~\ref{subsec:groups} we show that the luminosity evolution is not significant in the $0.3 < z < 0.9$ redshift interval. We also note that if we split sample C in two redshift bins, we obtain the same trends shown in Figure~\ref{badfrac} for both samples.

The 1-$\sigma$ uncertainties in the bad fractions were calculated from the confidence intervals (at confidence level, c$\approx$0.683) derived from binomial population proportions using the beta distribution \citep[see][for a description and justification of the method]{Cameron2011}.

It is evident that the fraction of kinematically-disturbed galaxies in clusters is  greater than in the field population, at least for $M_{B} < -20.5$. 
The percentage of ``bad'' over total number of galaxies in the whole $M_{B}$ and redshift range (of sample C) is 40$^{+5}_{-4}\%$ for clusters and groups ($44^{+6}_{-5}\%$ in clusters, $31^{+10}_{-7}\%$ in groups), and $25^{+4}_{-3}\%$ in the field.

It is important to recall that not all galaxies that were categorized as ``bad'' are necessarily kinematically disturbed, but the vast majority of them are. As explained in Section~\ref{subsec:quality}, some of them simply had poor quality spectra (e.g. badly subtracted sky lines near the studied galaxy emission line) but frequently, it is a difficult task to distinguish between these cases. Nonetheless, it is reasonable to argue that the results presented here are not biased because in principle, galaxies with bad spectra should appear in both field and cluster samples equally, and also their $M_{B}$ are distributed in the same way as the parent sample.
However, to verify that this is true, we examined all galaxy spectra again to make a very conservative cut that distinguishes kinematically-disturbed galaxies from the others (all the doubtful cases were rejected). We repeated the exercise presented in the left hand panel of Figure~\ref{badfrac} but this time, we only considered as ``bad'', those galaxies with clear and strong signs of kinematical disturbance in their spectra. By making these conservative cuts, the sample reduced to about half of its size. This is shown in the right hand panel of Figure~\ref{badfrac}, where we found similar trends as in the left hand panel, but for a smaller number of galaxies.  Numerically, the percentage of (confirmed) kinematically disturbed over total number of galaxies in the whole $M_{B}$ range is $22^{+6}_{-5} \%$ for clusters, $17^{+10}_{-5} \%$ for groups, $21^{+5}_{-4} \%$ for clusters and groups, and $13^{+3}_{-2} \%$ for field galaxies. Because of the difficulties in separating kinematically-disturbed galaxies from the rest, and having shown that the cut adopted does not bias the trends with magnitude and environment, we adopt the first cut (shown in the left panel of Figure~\ref{badfrac}) hereafter. 

Figure~\ref{badfrac} shows that, in clusters, the fraction of kinematically-disturbed galaxies is higher at brighter magnitudes. This does not happen in the field (or the effect is too mild to detect). It is not clear whether groups follow more closely the cluster or field behaviour (more detailed discussion in Section~\ref{subsec:environment}).
A possible interpretation is that the trend observed in clusters could be the result of fainter (less massive) cluster galaxies having already been stripped of their gas completely. This would cause them to have no (or very little) emission in their spectra, and are hence excluded from our emission-line galaxy sample. Nonetheless, it is arguable whether this could be a consequence of a larger fraction of early-type galaxies, (which are more likely to have disturbed rotation curves, as shown in Figure~\ref{bad_frac_morph}) at higher luminosities in clusters. We discarded this possibility by repeating the exercise shown in Figure~\ref{badfrac} with only the morphologically confirmed spirals. The results we obtain are compatible with our findings for the entire emission-line sample but are inevitably affected by larger uncertainties due to the reduced number of galaxies. 

In addition to the above interpretation, it is arguable that the most luminous galaxies are those that were accreted more recently and therefore our results reflect the influence of the cluster environment at play. In a hierarchical Universe, one expects more massive systems to be accreted later, although there is some scatter (De Lucia et al., in preparation). In Section~\ref{subsec:environment} however, we show that the fraction of kinematically-disturbed galaxies decreases with distance from the cluster centre (see Figure~\ref{r200}), hence the above interpretation is unlikely. 
The results of Section~\ref{subsec:environment} suggest another possibility: $f_{\rm K}$ may grow with
luminosity because brighter (emission-line) galaxies may be more likely to
reside in the cluster centres, where there is a higher incidence of
kinematically-disturbed galaxies. We discard this possibility since we find
no correlation between the luminosity of the cluster galaxies and their
distance to the cluster centre.

\subsection{Probing the Environment}
\label{subsec:environment}

\begin{figure}
\begin{center}
  \includegraphics[width=0.49\textwidth]{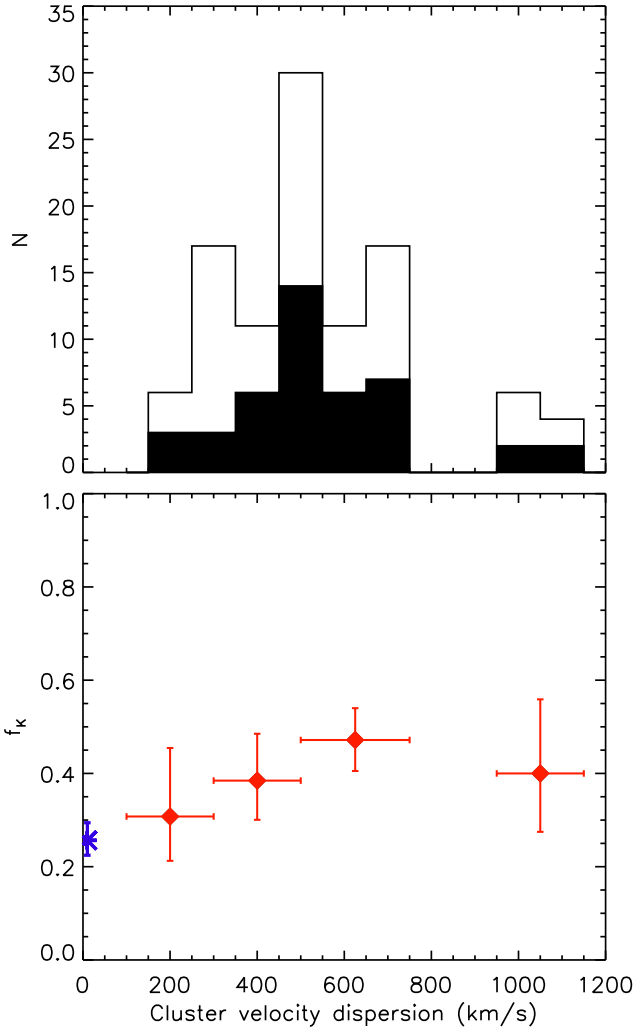}
\end{center} 
\caption{Top: The cluster velocity dispersion distribution of all galaxies in sample C (open histogram) and the distribution of those with ``bad'' fits (filled histogram) are plotted. Bottom: The fraction of ``bad'' galaxies (i.e. galaxies with disturbed kinematics) is shown as a function of cluster velocity dispersion for the matched sample C. The blue asterisk at $\sigma_{\rm cluster}\simeq0$ km/s corresponds to the field population, shown for comparison. The values of $\sigma_{\rm cluster}$ were taken from \citet{Halliday2004}, \citet{MJ2008}, and \citet{Poggianti2009}. A non-parametric Spearman's rank correlation coefficient analysis indicates that the correlation shown in this figure is significant at the 83\% level.}
 \label{cl_vel_disp}
\end{figure}

\begin{figure}
\begin{center}
  \includegraphics[width=0.49\textwidth]{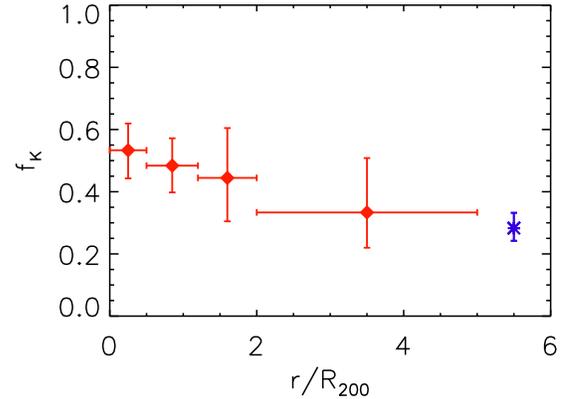}
\end{center} 
\caption{The fraction of galaxies with disturbed kinematics is shown as a function of $r/R_{200}$ for the luminosity-limited sample C. All the ``bad''  rotation curve fits are considered. The data point for the field is plotted for comparison at arbitrarily high $r/R_{200}$ in a blue asterisk. There seems to be significantly more galaxies with disturbed gas kinematics towards the cluster centre than in the field or high cluster-centric distances. A Spearman's rank correlation test indicates that the correlation shown in this figure is significant at the 98\% level.}
 \label{r200}
\end{figure}

\begin{figure}
\begin{center}
  \includegraphics[width=0.49\textwidth]{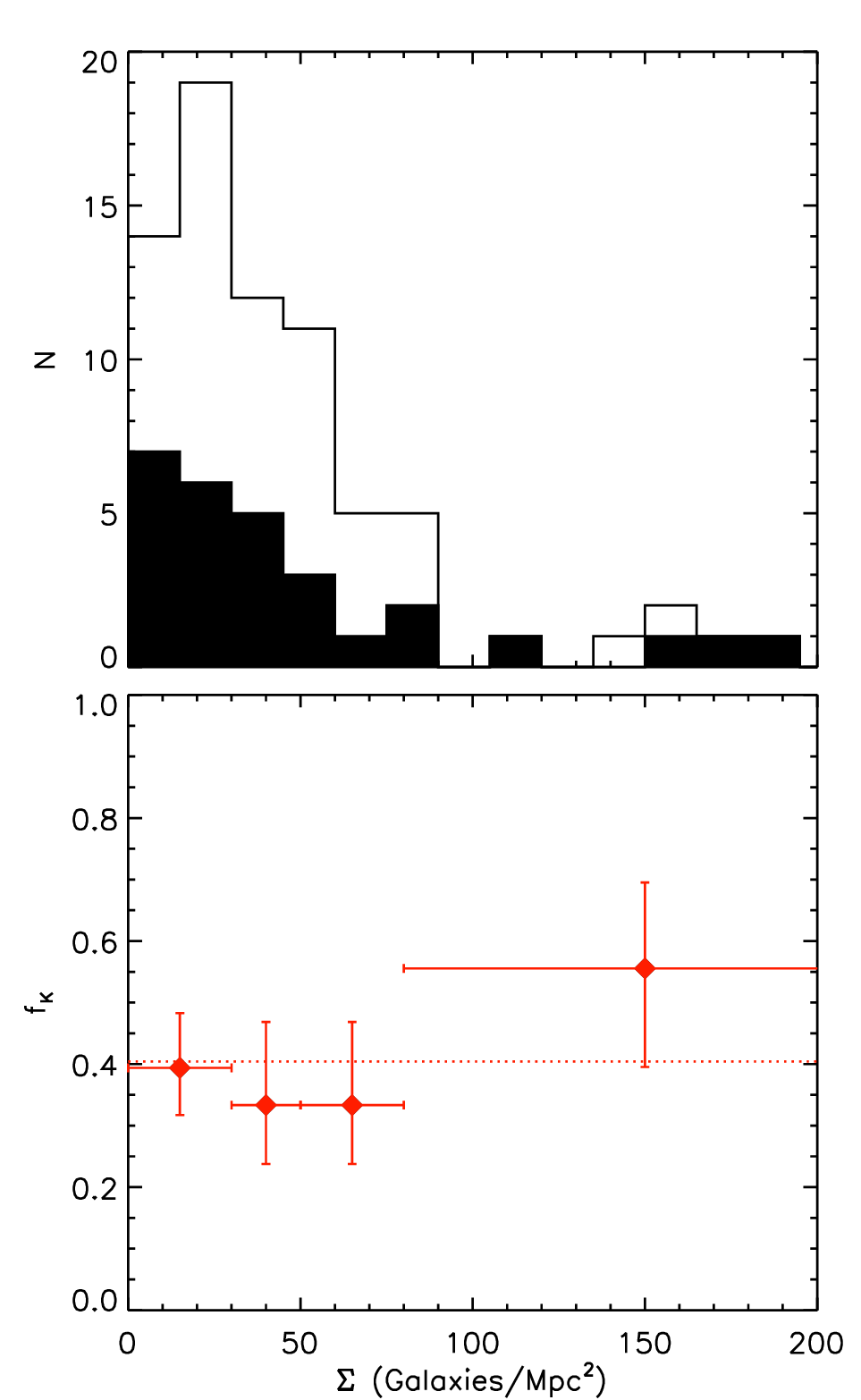}
\end{center} 
\caption{The top panel shows the distribution of the projected densities of the cluster emission-line galaxies in sample C. In the bottom panel, the fraction of galaxies with disturbed kinematics is shown as a function of projected density for cluster/group galaxies in the luminosity-limited sample C. The horizontal dotted line corresponds to the mean value of $f_{\rm K}$ and is plotted to show that the fraction of kinematically-disturbed galaxies is consistent with that value at all densities.}
 \label{dens}
\end{figure}

In Section~\ref{subsec:badfractions}, we compared the gas kinematics of cluster, group and field galaxies. There are other ways however of studying environmental effects on the galaxies' gas state.  In this Section, we investigate the dependence of the fraction of kinematically-disturbed galaxies with (i) velocity dispersion of the galaxies' host cluster, (ii) projected distance from the galaxy to the cluster centre, and (iii) projected galaxy density. 

A useful way to quantify the global environment in which a galaxy resides is in terms of the cluster velocity dispersion of the parent cluster ($\sigma_{\rm cluster}$),  a good proxy for the cluster mass. The top panel of Figure~\ref{cl_vel_disp} shows the cluster velocity dispersion distribution of all the cluster emission-line galaxies (open histogram) in the matched sample C, and highlights the distribution of galaxies with bad fits or kinematical disturbances (filled histogram). The cluster velocity dispersion range covered by EDisCS is very broad and thus is a good probe of environmental effects on galaxy properties. The bottom panel of Figure~\ref{cl_vel_disp} shows the fraction of kinematically-disturbed galaxies as a function of cluster velocity dispersion. This plot reinforces the results presented in Section~\ref{subsec:badfractions}, showing that the  fraction of kinematically-disturbed galaxies increases with $\sigma_{\rm cluster}$ by a factor of $\sim 1.5$ between $\sigma_{\rm cluster} \simeq 100 - 800$ km$/$s (followed by a point of significant uncertainty at $\sigma_{\rm cluster}\sim 1100$ km$/$s). A non-parametric Spearman's rank correlation coefficient analysis of the trend shown in Figure~\ref{cl_vel_disp} indicates that the observed correlation is only significant at the 83\% level, so this result needs to be confirmed with larger samples.

A frequently-used way of quantifying the local environment for a galaxy is the distance from the cluster centre, which should be correlated with, among other things, the density of the intracluster medium (ICM) and the velocities of the galaxies inside that radius. To compare galaxies in all clusters, we normalize the distance from the galaxy to the centre of the cluster ($r$) by  $R_{200}$, and study the ratio  $r/R_{200}$.
The values of $r/R_{200}$ used here were computed  in \citet{Poggianti2006}.
Figure~\ref{r200} shows the fraction of kinematically-disturbed galaxies as a function of $r/R_{200}$. The blue point corresponds to the field population and is plotted for reference at arbitrarily large radius.
The figure shows a clear trend of increasing disturbance towards the cluster centre. This correlation is also significant at the 98\% level.

We investigate how the fraction of kinematically-disturbed galaxies is affected by projected galaxy densities. The projected local galaxy densities used here are described in \citet{Poggianti2008}. Briefly, densities were computed  for each spectroscopically confirmed cluster member. They were derived from the circular area ($A$) that, in projection on the sky, encloses the $N$ closest galaxies brighter than an absolute $M_{V}$ limit. Hence, the projected density is $\Sigma=N/A$ and is given in number of galaxies per square megaparsec. The value of $N$ used was 10, and the limiting magnitude was $M_{V}=-20$. In this paper we use the density computed from the ``statistical subtraction method'' described in \citet{Poggianti2008}. In this method, all galaxies in the EDisCS photometric catalogues are used, and  $\Sigma$ is then corrected using a statistical background subtraction. We note that the calculations made in \citet{Poggianti2008} excluded two fields without deep spectroscopy, and two others that have a neighbouring rich structure at slightly different redshift, indistinguishable by photometric properties alone. For this reason, our $\Sigma$ analysis contains only part of our matched sample C, but this fraction is nonetheless significant.

Figure~\ref{dens} (bottom panel) shows the fraction of kinematically disturbed  cluster/group galaxies in the luminosity-limited sample C as a function of projected densities. 
It is clear that the fraction of kinematically-disturbed galaxies remains constant with $\Sigma$, up to the highest densities. 

To test that the trends seen in Figures~\ref{cl_vel_disp},~\ref{r200}, and~\ref{dens} are not dominated by the inclusion of elliptical and S0 galaxies (which we know are more likely to be disturbed, see Figure~\ref{bad_frac_morph}), we repeated each plot without the known E/S0s and obtained the same trends. In addition, we repeated them with only confirmed spirals. Because we only have visual (HST) morphologies for about half of the sample, the number of galaxies reduces significantly. The observed trends for the spiral galaxy sample remain unchanged, but inevitably suffer from greater uncertainty.

Because of the small number of galaxies in the bins of Figures~\ref{cl_vel_disp},~\ref{r200} and ~\ref{dens}, we adopted a conservative approach in estimating the confidence intervals \citep[the one described in][]{Cameron2011}. However, the clear and smooth trends that we observe in Figures~\ref{cl_vel_disp} and~\ref{r200} seem to suggest that we are overestimating the errors somewhat.

When comparing the results obtained from Figures~\ref{cl_vel_disp},~\ref{r200}, and~\ref{dens}, it is clear that the gas kinematics is not affected by the local galaxy density,  but significantly affected by the nature of the global environment itself (cluster mass and distance from centre). This strongly suggests that what affects most the properties of the gas in cluster galaxies has to be linked to the ICM and/or the gravitational potential of the cluster itself and not to galaxy interactions. 

\subsection{Morphologically Disturbed Galaxies}
\label{subsec:morph_badfractions}

With the aim of comparing the state and distribution of the gas and the stars for galaxies in different environments, we performed an independent analysis of the morphological disturbances of the galaxies, as traced by optical (HST) imaging. The expectation
is that our analysis of the 2D spectroscopy we have just described provides information on the gas structure and distribution, while the
optical light traces the stellar structure. For the 155 (out of 264) galaxies with HST observations in the  luminosity limited sample C, we fitted a smooth single-Sersic index model. We used the GALFIT code, described in \citet{Peng2002}. The set-up with which GALFIT runs, named GALPHYT\footnote{Developed in python by Carlos Hoyos}, is described in detail in \citet{Hoyos2011}.
Residual images were created by subtracting the model from the galaxy's HST image. These residuals highlight the presence of morphological distortions and contain valuable information about the interaction state.

\begin{figure}
\begin{center}
  \includegraphics[width=0.49\textwidth]{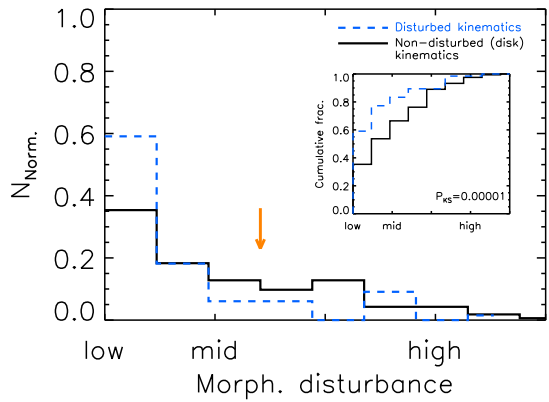}
\end{center} 
\caption{The distribution of the degree of morphological distortion is plotted in a histogram with normalized area (to unity) for: galaxies with good rotation curve fits (i.e. with normal disk kinematics, shown in a black solid line), and galaxies with disturbed disk kinematics (blue dashed line). The vertical (orange) arrow indicates the limit where we have separated non-disturbed from disturbed morphologies (definition used for Figure~\ref{morphbadfrac}). The sample plotted is the luminosity limited sample C that counts with HST observations (see circled symbols in Figure~\ref{matched_sample}). The inset panel shows the cumulative distributions of the morphological disturbance, as well as KS statistics, for the kinematically disturbed and undisturbed galaxies.}
 \label{asy}
\end{figure}

\begin{figure}
\begin{center}
  \includegraphics[width=0.5\textwidth]{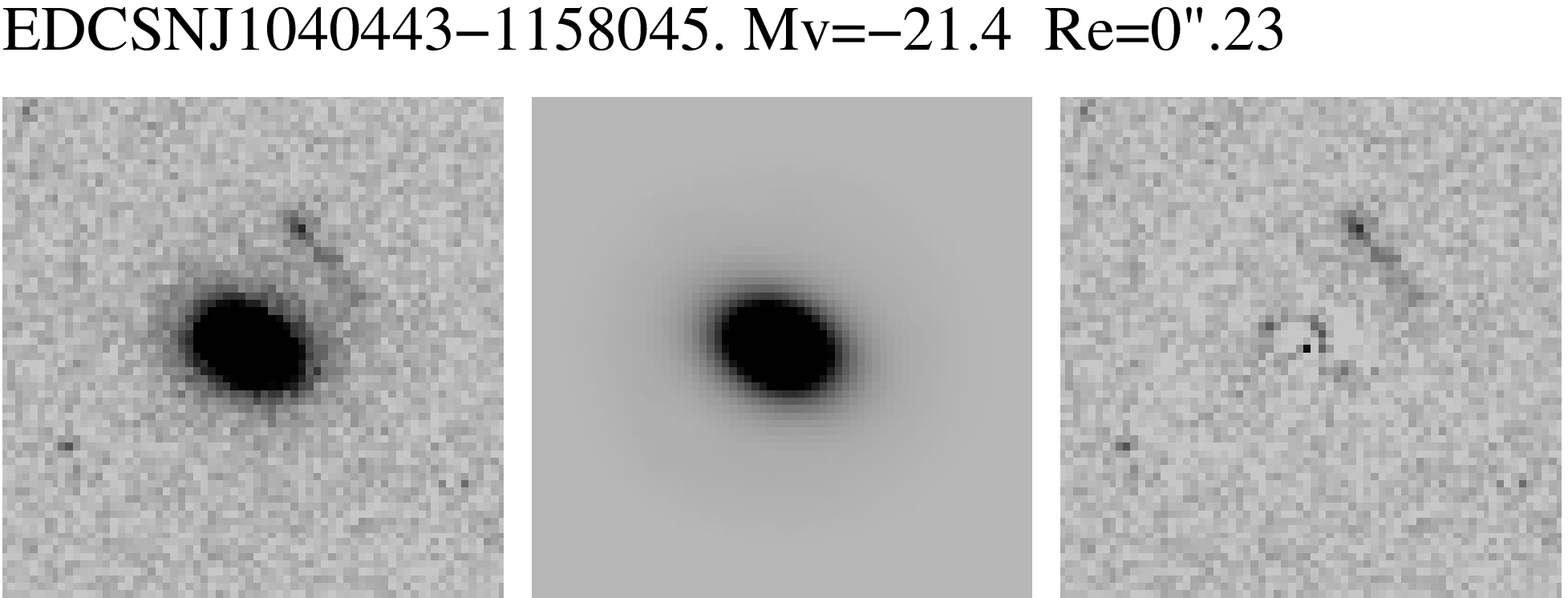}
  \includegraphics[width=0.5\textwidth]{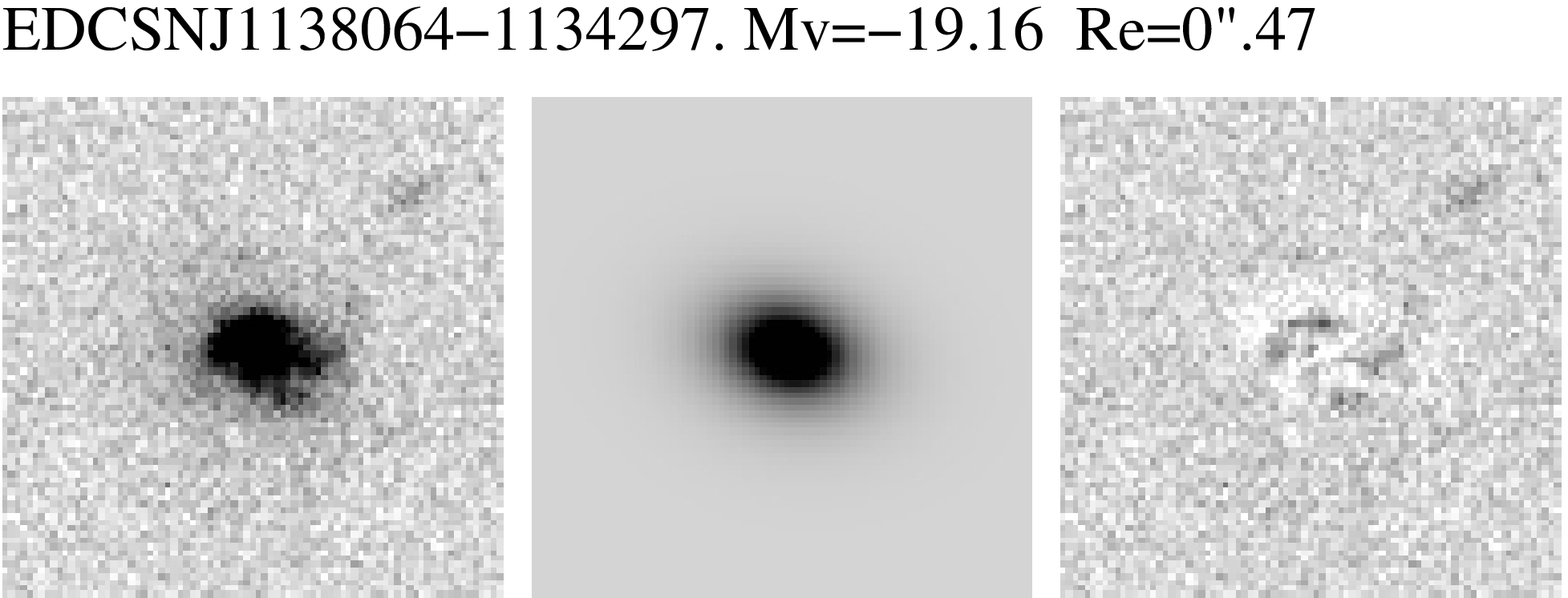}
  \includegraphics[width=0.5\textwidth]{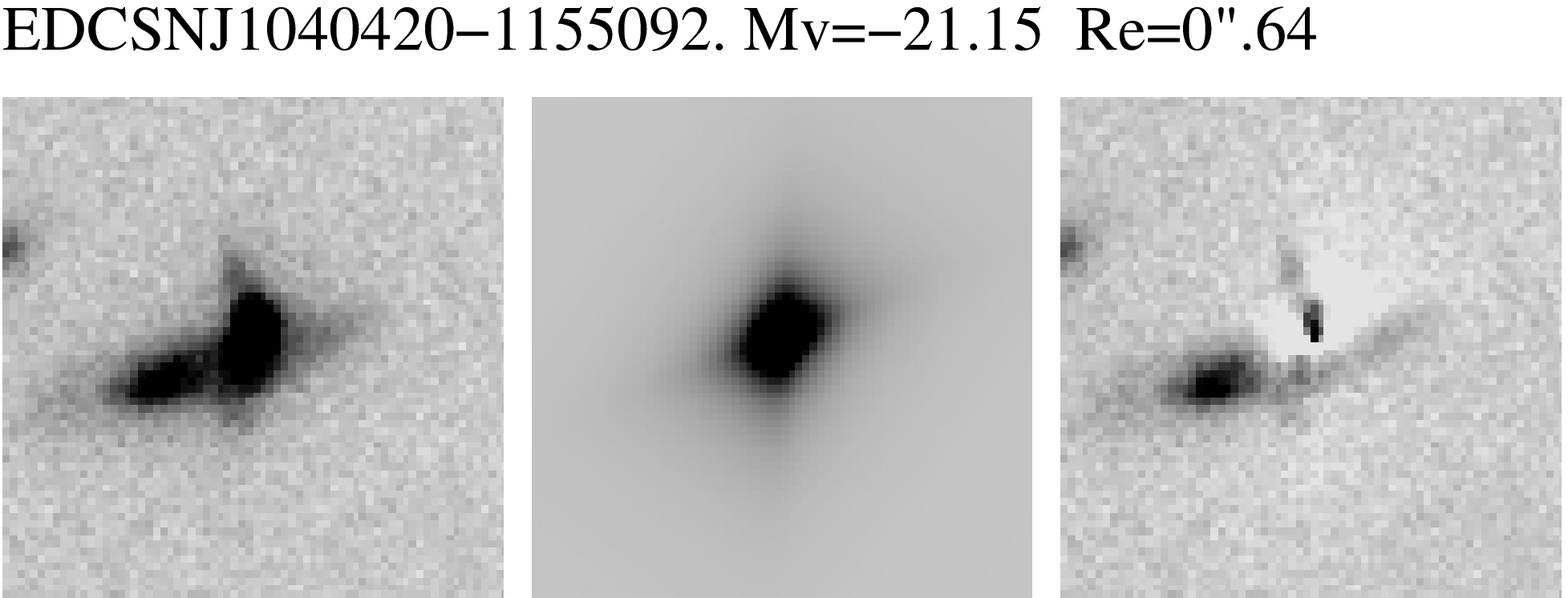}
  \includegraphics[width=0.5\textwidth]{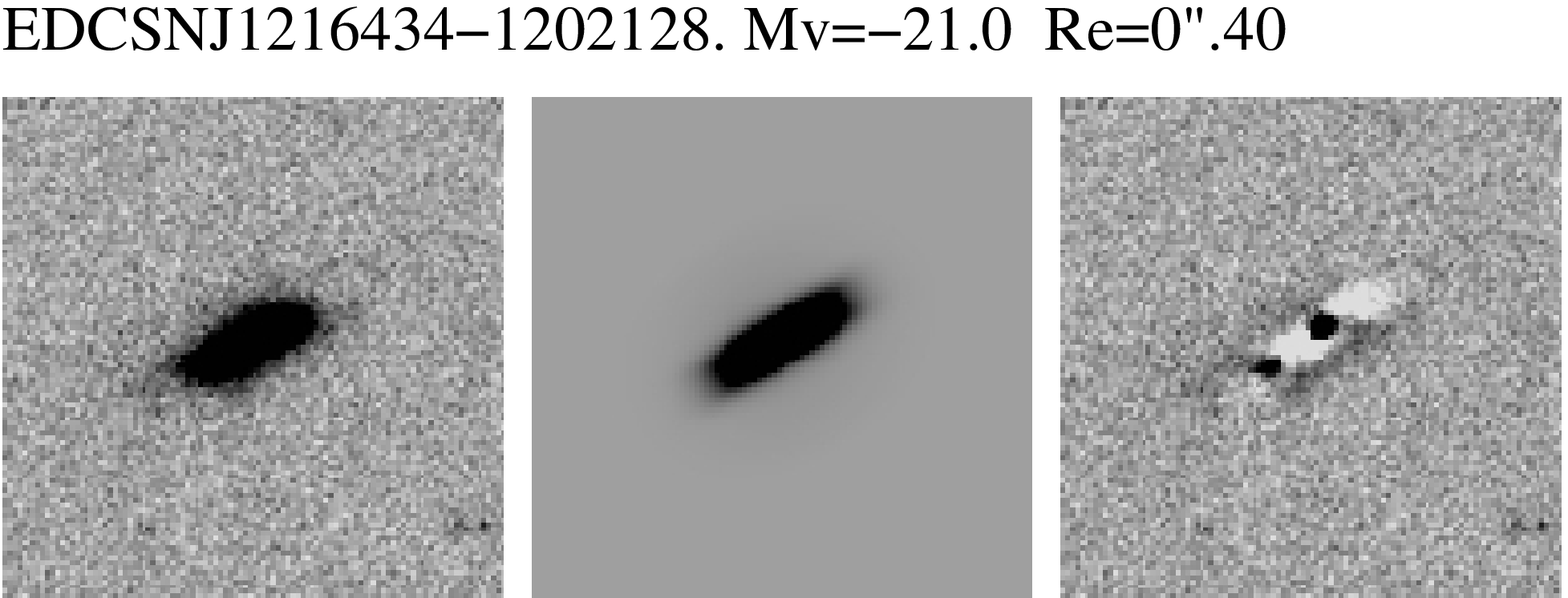}
\end{center} 
\caption{Representative examples of the method used to identify morphological disturbances in our galaxy sample. Our results are shown for four galaxies, the two on the top were considered ``good fits'' or morphologically undisturbed galaxies, while the other two were classified as ``morphologically disturbed''. The first column presents the HST cutout of the galaxies, the second column shows the best-fit model made to that image, and the third column exhibits the residual image between the model and the data. Galaxy names, $M_V$ and effective radius are listed at the top of each galaxy.}
 \label{sersic_eg}
\end{figure}

Three of the authors of this paper (AAS, CH, and YLJ) independently examined the residual images and graded the level of morphological disturbance of the galaxies under study. We did this by looking for different features such as asymmetry, presence of tidal tails, nuclear components, mergers, and interactions. Each of these parameters were graded separately. By comparing the parameter space drawn by each examiner, we reached the conclusion that the most reliable (and consistent) way of determining the degree of morphological disturbance was the quantification of the asymmetry in the residual image. Hence, we defined a morphological disturbance index by combining the  grades for the asymmetry parameter from the different examiners into an average grade.  The disturbance index increases from 0 in the positive direction as the level of asymmetry becomes stronger. From the distribution of morphological distortion  in our sample, we then defined two sub-samples of morphologically ``good'' and morphologically disturbed galaxies by choosing a threshold value (see orange arrow in Figure~\ref{asy}). Figure~\ref{sersic_eg} shows a few examples of what we call morphologically disturbed and undisturbed galaxies\footnote{The complete set of HST images, single-Sersic fits and residuals images for the EDisCS galaxies treated in this paper can be found in the EDisCS website at: http://www.mpa-garching.mpg.de/ediscs/Papers/Jaffe\_tfr\_2011/single\_sersic\_fits.html}.

%http://www.nottingham.ac.uk/$\sim$ppxyj/single\_sersic\_fits.html}. 

To understand how the morphological and kinematical disturbances are related, we compared the morphological disturbance index for both the kinematically ``good'' and ``bad'' samples. This is illustrated in Figure~\ref{asy}, where we have plotted the morphological distortion index for the galaxies with  disturbed gas kinematics (dashed blue line) and the galaxies with good rotation curve fits (solid black line). The figure also contains an inner plot showing the cumulative distributions and the resulting KS statistics.
We find that although the distributions are statistically different ($P_{\rm KS}=10^{-5}$), there does not seem to be a very clear difference between the morphological disturbance indices of galaxies with perturbed and unperturbed gas distributions. This suggests that the disturbance we observe in the gas is not directly linked to the galaxy’s morphological distortions.

Keeping in mind that early type galaxies are more likely to be kinematically disturbed (see Figure~\ref{bad_frac_morph}), and that there are more kinematically-disturbed galaxies in clusters than in the field (see Figure~\ref{badfrac}), we repeated the analysis shown in Figure~\ref{asy} with only morphologically-classified spiral galaxies, separating cluster and field ones. The results did not change significantly.

\begin{figure}
\begin{center}
 \includegraphics[width=0.49\textwidth]{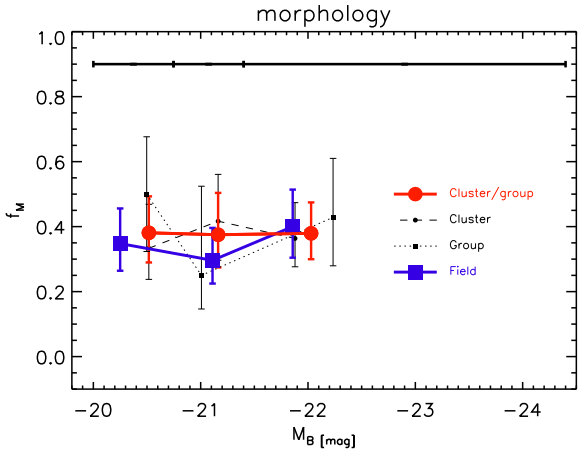}
\end{center} 
\caption{Fraction of morphologically disturbed galaxies in different environments as a function of $M_{B}$, for galaxies with HST data in the matched sample C. This plot is analogous to the ones shown in Figure~\ref{badfrac}, but instead of showing the disturbance in the gas kinematics with environment, it studies the disturbances in the stellar structure. The different symbols correspond to different environments, as shown in the legend. The error bars and $M_{B}$ bins (shown at the top of the plot) are as in  Figure~\ref{badfrac}. We observe no dependence of morphological disturbance on environment.}
 \label{morphbadfrac}
\end{figure}

We also studied the fraction of morphologically disturbed galaxies, $f_{\rm M}$, as a function of $M_{B}$ in the same manner of Section~\ref{subsec:badfractions}. The result is shown in  Figure~\ref{morphbadfrac} (plotted in the same way as Figure~\ref{badfrac} for comparison). We observe that there is no significant difference between the morphologically disturbed galaxy fraction between cluster, group, and field environments in the $M_B$ range studied. Our results are actually consistent with a constant morphologically disturbed fraction as a function of $M_{B}$ in all environments.
The total fraction of  morphologically disturbed galaxies (over the full $M_B$ and redshift range of sample C) is 
$ 47\pm 7$\%  in clusters, $41^{+12}_{-10}$\% in groups, $45\pm 6$\% in cluster and groups, and $49\pm6$\% in the field. It is important to point out that these fractions should only be compared internally within our study since the actual value of $f_{\rm M}$ will depend on the definition of ``kinematically disturbed'' or ``morphologically disturbed''. For instance, if we shift the vertical arrow in Figure~\ref{asy} that defines the threshold between kinematical disturbed and non-disturbed galaxies, the fractions change in number. However, the lack of a trend seen in  Figure~\ref{morphbadfrac} does not change. We emphasize that the high fraction of disturbed galaxies (of $\sim50$\%, cf. Figure~\ref{morphbadfrac}) is a direct result of the threshold used to define morphological disturbances (the orange arrow in Figure~\ref{asy} roughly divides the galaxy sample in half). Moreover, by subtracting a smooth model to the HST images, small morphological disturbances are enhanced \citep[cf.][]{Hoyos2011}, increasing the number of galaxies categorized as ``morphologically disturbed''.

We note that, our index for morphological disturbance (a visual index) is very similar to the asymmetry index in the CAS system \citep{Conselice03}. Our threshold value for defining morphologically disturbed galaxies is approximately equivalent to a CAS
asymmetry index greater than 0.2. 
When using CAS asymmetry measurements and adopting this threshold value, we obtain the same trends observed in Figures~\ref{asy} and \ref{morphbadfrac}.

The results presented here for the disturbance of the structure of the galaxies' stellar component and those from Section~\ref{subsec:badfractions}, show that the fraction of kinematically-disturbed galaxies is higher in clusters, whilst the fraction of morphologically disturbed galaxies does not change significantly with environment (see Figures~\ref{badfrac} and ~\ref{morphbadfrac}). This suggests that environmental effects are mild enough to not disturb the stellar structure in the galaxies significantly, but to strongly affect the gas in cluster environments. The implications of this result will be discussed in Section~\ref{sec:discusion}.

\subsection{The Tully-Fisher Relation of cluster and field galaxies}
\label{subsec:TF}

One of our principal aims is to study possible variations with environment of the TFR  to help us understand what happens when field galaxies fall into a cluster. Having created matched cluster and field galaxy samples (Section~\ref{sec:matched}), we proceed to construct Tully-Fisher diagrams and compare the distribution of cluster and field galaxies in them. For this study, we only use galaxies with good rotation-curve fits. To ensure these galaxies are supported by rotation, we checked that their computed velocities were consistent with non-zero rotation by rejecting galaxies with $V_{\rm rot} < 2 \sigma^{-}_{V_{\rm rot}}$, where $\sigma^{-}_{V_{\rm rot}}$ is the left-hand side error on $V_{\rm rot}$. Forty-five of our ``good'' galaxies were consistent with no rotation. Typically, these galaxies  have $V_{\rm rot} \sim 15$ km/s and $\sigma^{-}_{V_{\rm rot}} \sim 20$ km/s.  Their morphology distribution is as broad as the parent sample, with a higher number of irregular galaxies, and their $M_B$ mimics the sample of ``good'' galaxies, peaking at $\sim-20.3$ mag.

\begin{figure*}
\begin{center}
  \includegraphics[width=0.8\textwidth]{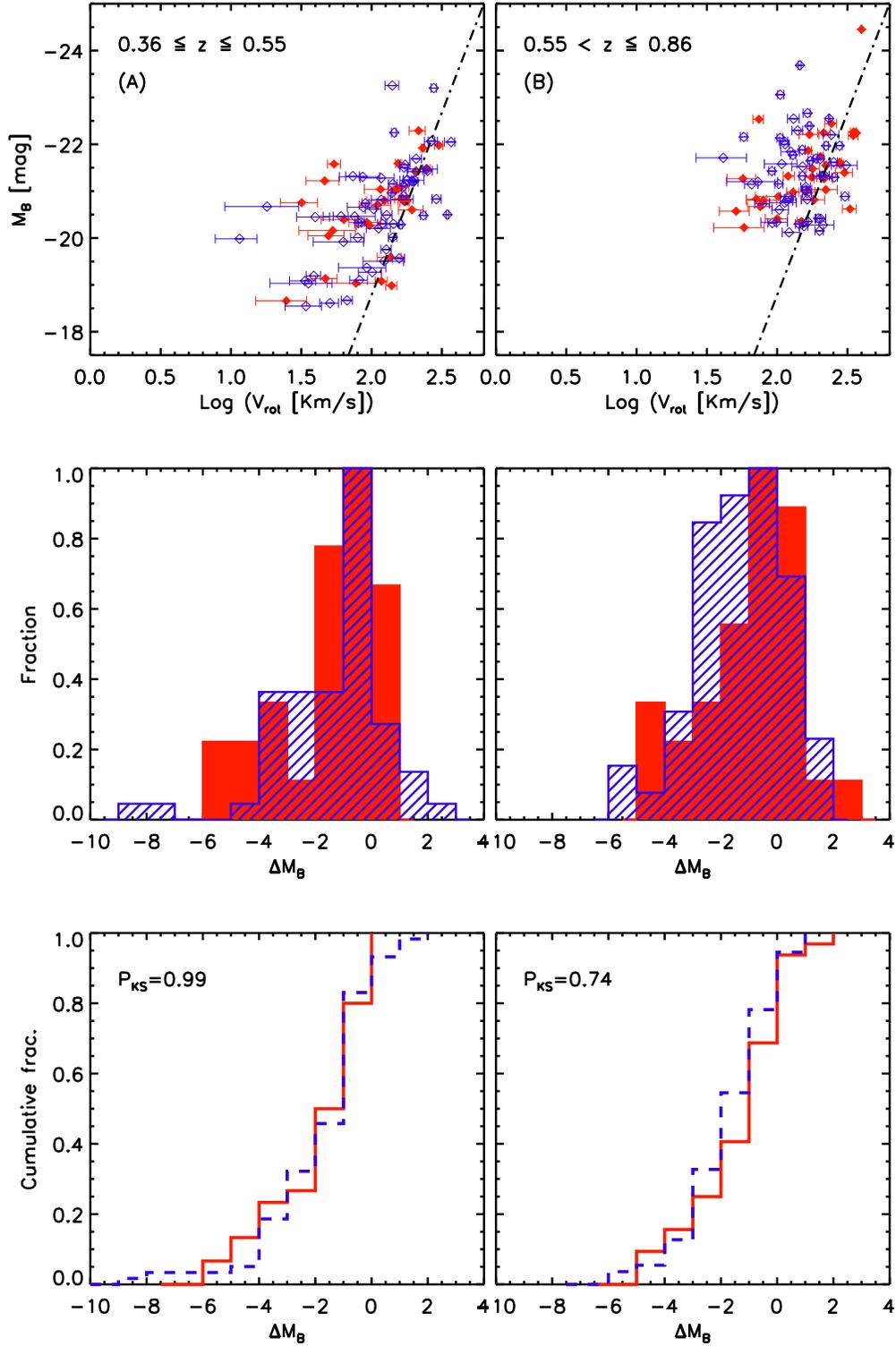}
\end{center} 
\caption{$M_{B}$ vs. $\log V_{\rm rot}$ is plotted in the two upper panels for the low and mid-z samples (A and B, respectively, as labeled). As in Figure \ref{matched_sample}, the cluster/group galaxies are plotted as red filled diamonds, and the matched field sample corresponds to the blue open diamonds. The fiducial TFR of \citet{Tully1998} is marked by the dotted-dashed line in both panels. The middle panels show the distribution of the vertical difference between the points and the plotted line ($\Delta M_{B}$) for cluster/group (red) and field (blue) galaxies for each sub-sample. The bottom panels show the cumulative distributions of $\Delta M_{B}$ in each case. The Kolmogorov-Smirnov (KS) probability that the two samples follow the same distribution, $P_{\rm KS}$, is shown in a corner.}
 \label{TFR}
\end{figure*}

The top panels in Figure \ref{TFR} shows our TFRs. The absolute rest frame $B$-magnitude is plotted against the $\log V_{\rm rot}$ for cluster/group galaxies (red symbols) and field galaxies (blue symbols) for the low and mid-z matched samples (sample A in the left hand panel and sample B in the right hand panel). The fiducial local TFR of \citet[][from now on T98]{Tully1998} is shown as a dotted-dashed line in both panels for reference. A relation can be seen in both samples, although the $M_{B}$ limit of sample B confines the TFR to a range of a few magnitudes. The observed  scatter in the TFR is $0.233$ dex in $V_{\rm rot}$. This scatter is not dominated by the errors in $V_{\rm rot}$, which are typically $\sim 0.07$ dex. The intrinsic scatter we measure is thus 0.230 dex. Our scatter is larger than local studies of the TFR but smaller than similar studies at high redshift. For example, the TFR presented here has lower scatter than that of  \citet{Kassin2007}. As mentioned above, they are able to reduce it significantly by replacing rotation velocity with a kinematic estimator, which accounts for non-circular motions through the gas velocity dispersion. In this paper, owing to our poor spectral resolution, we are unable to measure velocity dispersions, and hence apply their method. In Section~\ref{subsec:S_TFR} however, we  show that the scatter is reduced if we limit our sample to spiral galaxies only.

To compare the cluster and field TFRs, we define the quantity $\Delta M_{B}$ as the vertical difference in $M_{B}$ between our data points and the local relation plotted.
The middle and bottom panels in Figure~\ref{TFR} show the $\Delta M_{B}$ distributions and cumulative distributions respectively, again for the two redshift ranges of our A and B matched samples. The fact that the $\Delta M_{B}$ distribution peaks at $\sim-1\,$mag is probably due to some evolution with redshift of the TFR \citep{Vogt96,Bamford05,Bamford06,Weiner2006}. However, since it is
extremely difficult to make direct reliable comparisons between TFRs at $z\sim0$ and at intermediate $z$ \citep[see, e.g.][]{Weiner2006}, we will not attempt to quantify this evolution here and only make comparisons internally within our sample for which the selection effects and measurement biases are the same.
From these plots, we can see that cluster/group and field galaxies have a remarkably similar distributions of $\Delta M_{B}$, implying that they follow the same TFR. When applying a KS test to the matched sample A (left hand panels), we obtained a probability that the 2 samples are drawn from the same distribution of $P_{\rm KS}=0.99$. 
In sample B (right hand panels), $P_{\rm KS}=0.74$. These numbers are also shown in the bottom panel of Figure~\ref{TFR}.

Although no difference is observed between cluster/group and field TFR in $M_B$ (for a fixed  $V_{\rm rot}$), it is still possible that a difference could arise in their $V_{\rm rot}$ for a fixed $M_B$. To test whether this hypothesis is true, we computed the horizontal (velocity) difference between the data points and the local TFR ($\Delta V_{\rm rot}$). Again, no difference between cluster/group and field galaxies is observed.  

The lack of evidence for environmental effects on the TFR could be caused by the fact that we cannot plot the kinematically-disturbed galaxies on our Tully-Fisher diagrams, as their rotational velocities cannot be reliably measured. 
If there were an enhancement/suppression of the star formation in galaxies falling into clusters (hence an increased $B$-band luminosity), this should be more easily seen in the galaxies that already show signs of gas disturbance. However, it is precisely these galaxies (flagged as kinematically ``bad'')  we rejected because of our inability to fit a robust rotation curve (from which we could measure $V_{\rm rot}$). 
Nevertheless, if we take the observed lack of differences between the field and cluster TFRs at face value, we would conclude that there is no significant enhancement in the star formation of the infalling galaxies (which presumably could have been caused by environmental effects such as mergers in the cluster outskirts or compression of the interstellar medium by interaction with the clusters' dense intergalactic medium). 
However, it is clear that additional independent evidence is needed to draw definitive conclusions. To achieve this we will combine the TFR results shown here with a study of the star formation activity of the galaxies in Section~\ref{subsec:ssfr}.

The lack of significant differences between the TFRs of field and cluster galaxies that we find here agrees with the work of \citet{Nakamura2006} and \citet{ziegler03}, but disagrees with the $3\sigma$ difference found by \citet{Bamford05}. While Nakamura et al. and Ziegler et al. carried out rotation curve quality controls similar to the ones performed here, Bamford et al. accepted fits of lower quality. To test whether this is the cause of the discrepant results we repeated our TFR analysis accepting the Vrot values derived from all the fits, including those from bad quality ones. We find that even when including the ``bad'' fits, we found no significant difference between the TFRs of cluster/groups and the field. We thus conclude that differences in the quality of the accepted fits are not responsible for the discrepant results obtained by Bamford et al. and ourselves. We offer no convincing explanation for this discrepancy,
but since our sample is significantly larger than theirs and the quality of our data is at least as good (and often better), we trust that our result is more robust.

\subsection{The difference between cluster and group galaxies in the TFR}
\label{subsec:groups}

Cluster cores can have severe effects on galaxies residing near it. Galaxies however, are thought to interact with harsh environments well before reaching the centre of a cluster. \citep{Kodama2001,Treu2003}. In the hierarchical scerario of structure formation, infalling groups of galaxies build the rich galaxy clusters we observe today. Galaxy groups are thus likely to represent a natural environment for galaxy \textit{preprocessing} \citep[e.g.][]{Fujita2004} through tidal interactions that would not be as effective in higher velocity dispersion environments.

In this section, we distinguish galaxies in clusters and groups in the quest for evidence of more refined environmental effects. We compare galaxies in clusters, groups, and the field with each other in the Tully-Fisher diagram in a similar manner to Section~\ref{subsec:TF}. 

\begin{figure}
\begin{center}
  \includegraphics[width=0.5\textwidth]{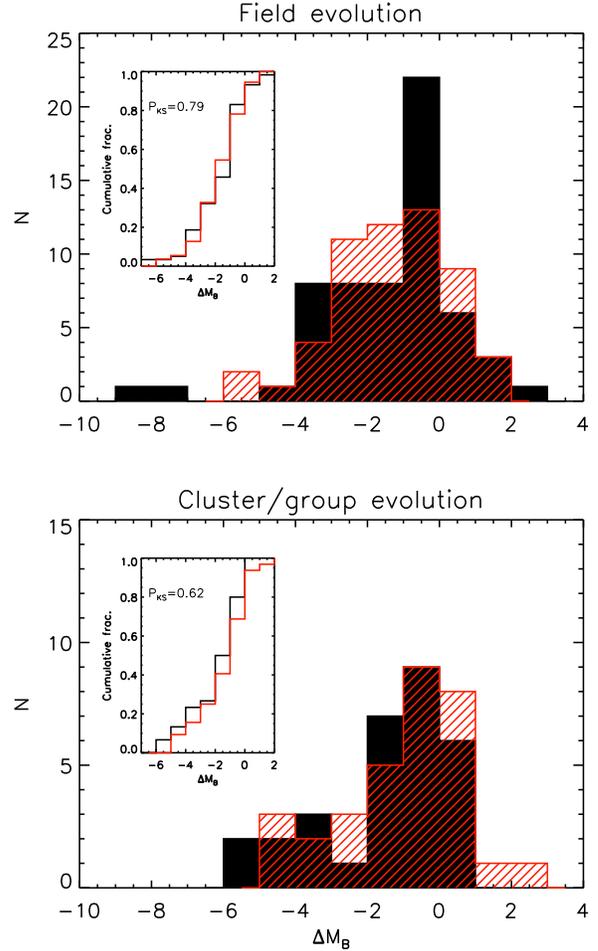}
\end{center} 
\caption{The distribution of $\Delta M_{B}$ for the field galaxies (upper panel) and cluster/group galaxies (lower panel). The black solid histogram in both panels corresponds to the lower redshift galaxies in the matched sample A, while the red, dashed histogram traces the higher redshift matched sample B (see Section~\ref{sec:matched} for the definition of these samples). In addition, each panel shows a smaller inner plot containing the cumulative distributions of samples  \textbf{A} and \textbf{B} for each case. These smaller plots also show the resulting KS statistics.}
 \label{evol}
\end{figure}

When distinguishing group from cluster galaxies our number counts inevitably drop. We therefore consider in this section the matched sample C that spans the redshift range $0.36\leqslant z \leqslant0.86$ and is limited by $M_B=-20$ mag. In this way, we improve the quality of our statistics. Because the redshift range of the full matched sample C is large, we first test whether evolutionary effects would bias this study in the following. We do not attempt however to perform an accurate TFR evolution study since it is very difficult to properly fit a TFR to high redshift galaxy samples (given the magnitude cuts and the amount of scatter present). For this reason, we only quantify evolutionary trends by comparing our data points with the local TFR, assuming the slope is constant across the entire redshift range.

The middle panels of Figure~\ref{TFR} showed that our matched samples have a brighter TFR than the local relation. We represent this with the quantity $\Delta M_B$, which equals the vertical difference between the galaxy's $M_{B}$ and the local TFR.
By comparing the same galaxy population (e.g. only field galaxies) in sub-samples A and B (at low and mid-z, respectively) against the local relation, we are able to quantify the TFR evolution from $z=0$ to the mean redshifts of samples A and B. 
The field galaxies of sample A, show a median $\Delta_{M_{B}}$ of  $= -0.93$ mag ($\langle \Delta M_B \rangle=-1.39$ mag), while, in the higher redshift sample B, they show  median $\Delta_{M_{B}}= -1.34$ mag  ($\langle \Delta M_B \rangle=-1.35$ mag). We emphasize that we do not attempt to make a detailed study of the TFR evolution here. Formally, this simple test suggests that there is a $\sim 1\,$mag evolution in the TFR's $M_B$, from $z=0$ to $z\sim 0.5$, in agreement with previous studies \citep{Vogt96,Bamford05,Bamford06,Weiner2006}.

We then looked for any evidence for evolution in $M_B$  in the range $0.36\leqslant z\leqslant0.86$ by comparing sub-samples A and B against each other. We did this separately for the field and cluster/group populations. Figure~\ref{evol} shows the $\Delta M_{B}$ distribution for the field (upper panel) and the cluster/group galaxies (lower panel). The black solid histogram corresponds to the lower redshift galaxies in sample A, while the red, dashed histogram traces the higher redshift sample B. In each panel, a smaller inner plot shows the cumulative distributions of samples  A and B, in addition to the KS statistics. From these plots, we see that although there is a significant offset in $M_{B}$ from the local relation, there is no evident evolution within the redshift range of our matched sample. In other words, we find weak or no evolution of the TFR in either field or cluster/group galaxies at $0.36\leqslant z\leqslant0.86$. 
This result allows us to compare different galaxy populations (cluster, group, and field galaxies) across the full redshift range of the matched sample C expecting redshift-dependent effects to be small.

\begin{figure*}
\begin{center}
  \includegraphics[scale=0.7]{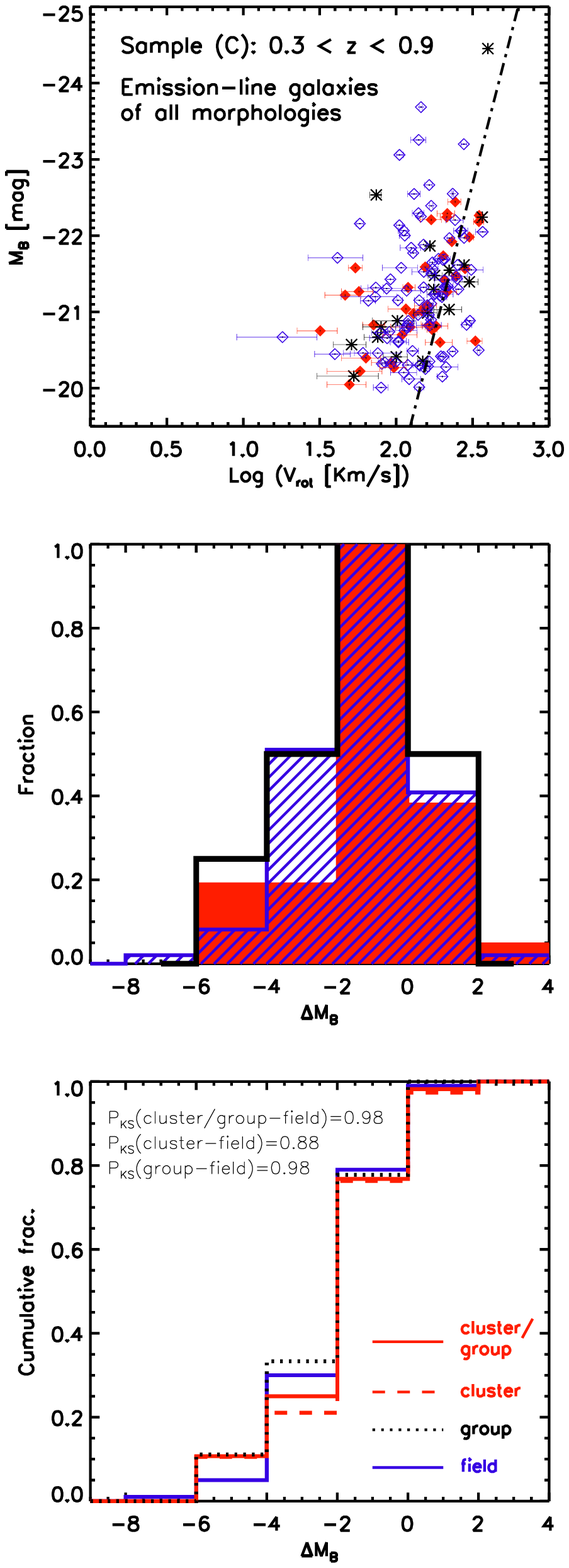}
  \includegraphics[scale=0.7]{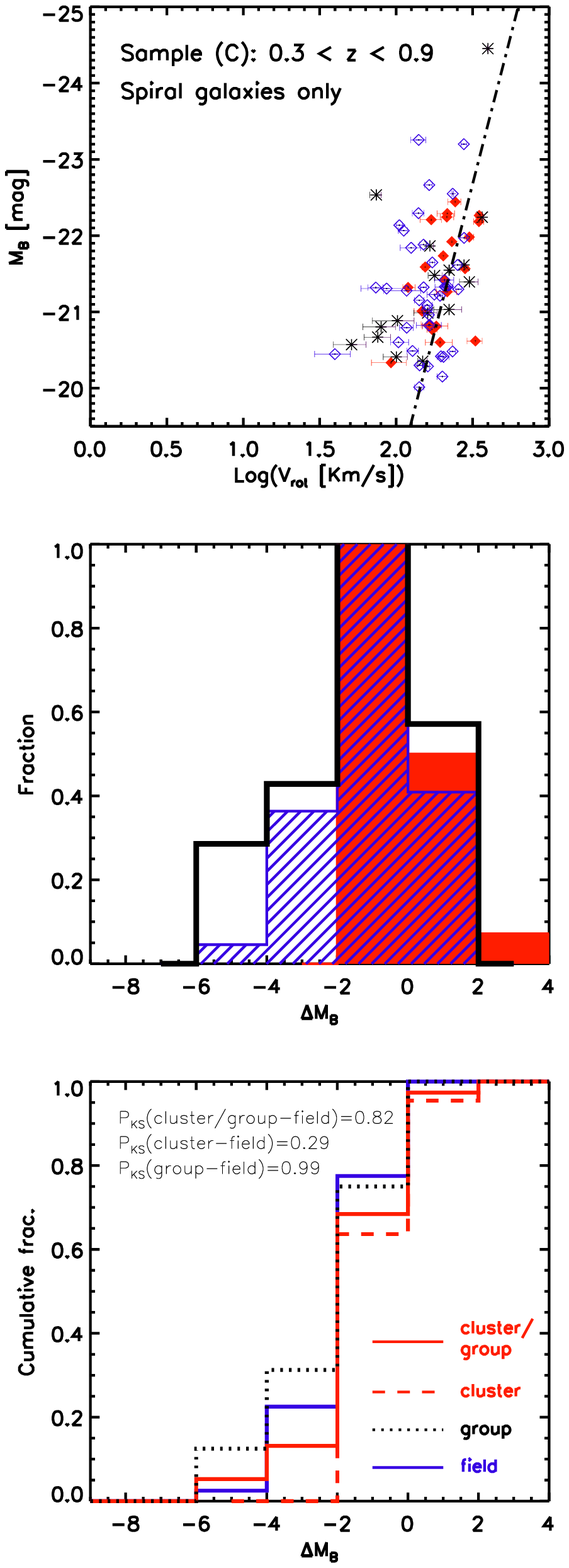}
\end{center} 
\caption{As in Figure~\ref{TFR}, $M_{B}$ vs.  $\log V_{\rm rot}$ for the galaxies in the matched sample C are plotted in the upper panel. Cluster galaxies are plotted as red filled diamonds, groups are represented as black asterisks, and the matched field sample corresponds to the blue open diamonds. The fiducial TFR of \citet{Tully1998} is marked by the dotted-dashed line. The middle panel shows the $\Delta M_{B}$ distribution for cluster (red, solid), groups (black, open), and field (blue, shaded) galaxies. The bottom panels show the cumulative distributions of $\Delta M_{B}$ for cluster (solid red line), group (dotted black line), and field (dashed blue line) galaxies. KS statistics are shown in the left hand side of the plot. The \textbf{left} hand panels consider all emission-line galaxies in sample C, whilst in the \textbf{right} hand panel, only morphologically classified spirals are plotted.}
 \label{groups}
\end{figure*}

The left hand side of Figure~\ref{groups} shows the absolute rest-frame $B$-magnitude plotted against $\log V_{\rm rot}$ for sample C. As in Figure~\ref{TFR}, the fiducial local TFR is again plotted (dotted-dashed line) for reference. The middle panel presents histograms of $\Delta M_B$  for cluster (solid red), group (open black), and field (dashed blue), while the bottom panel contains the cumulative distributions of $\Delta M_B$ for cluster (solid red line), group (dotted black line), and field (dashed blue line) galaxies. In addition, KS statistics are  shown in the left hand side of this plot.

We find that by making the distinction between group and cluster galaxies in the TFR, no significant differences arise. This can also be seen in the lower-left panel of Figure~\ref{groups}, where the cumulative fractions and KS statistics are shown. 
We still find no significant differences, suggesting again a lack of environmental effects on the TFR, at least when selecting emission-line galaxies that are not kinematically disturbed.

\subsection{The TFR of morphologically classified spirals}
\label{subsec:S_TFR}

It is well known that the TFR scatter is related to galaxy morphology \citep[e.g.][]{Kannappan2002} and it is arguable wether S0 and spiral galaxies, for example, should follow the same relation. Recent studies \citep{Bedregal2006,Williams2010} showed, that S0 galaxies have the same TFR slope as the spirals, but are on average fainter at a given rotational velocity. 

The TFR sample we have studied so far contains galaxies with unknown morphology and a few known not to be spirals. 
To study the effect of environment on the spiral-TFR, we extract the morphologically classified spirals from our matched sample C to construct a TFR of spirals only. Out of the 154 ``good''  emission-line galaxies in this sample (91 of which have HST observations, see circled symbols in Figure~\ref{matched_sample}), only 66 have a confirmed HST spiral morphology and velocities consistent with non-zero rotation. The top-right panel of Figure~\ref{groups} shows the spiral TFR at $0.3<z<0.9$. The distribution of galaxies in the TFR is tighter than that seen when plotting  all the emission-line galaxy sample (left hand side of Figure~\ref{groups}). The intrinsic scatter in the spiral-TFR is $0.18$ dex in $\log V_{\rm rot}$ (compared with $0.23$ dex if we consider all emission-line galaxies in the luminosity-limited sample). When comparing the distributions of the TFR residuals for the emission-line sample (left hand side of Figure~\ref{groups}) and morphologically classified spirals (right hand side of the Figure) for each environment, we find that the  distributions of group and field galaxies are remarkably similar, whilst the cluster galaxies show some deviation. In numbers, we obtained the following KS probabilities: 
$P_{\rm KS}=0.23$ for cluster members, 
$P_{\rm KS}=1.00$ for galaxies in groups, and 
$P_{\rm KS}=0.82$ in the field sample.

When studying the environmental effects on the spiral-TFR, we again observe no difference between the TFR residuals of field and cluster/group galaxies (see solid blue and solid red lines in the bottom-right panel of Figure~\ref{groups}), but this time, a small difference between cluster ($\sigma_{cl}>400$ km$/$s structures; dashed, red line) and field galaxies seems to appear. However, its significance is too small ($P_{\rm KS}=0.29$, see cumulative fractions and KS statistics in the bottom-right panel of Figure~\ref{groups}) to consider it too seriously. When combining cluster and group galaxies into one (more numerous) sample, and comparing with the field, this difference becomes negligible ($P_{\rm KS}=0.82$).

Complementary to the results found in this section, and in Sections ~\ref{subsec:groups} and~\ref{subsec:TF}, we investigated possible correlations between TFR residuals ($\Delta$M$_B$) with cluster velocity dispersion, distance from the cluster centre and projected galaxy density, and found that there are no obvious trends with environment.

\subsection{Star formation}
\label{subsec:ssfr}
 
\begin{figure}
\begin{center}
  \includegraphics[width=0.49\textwidth]{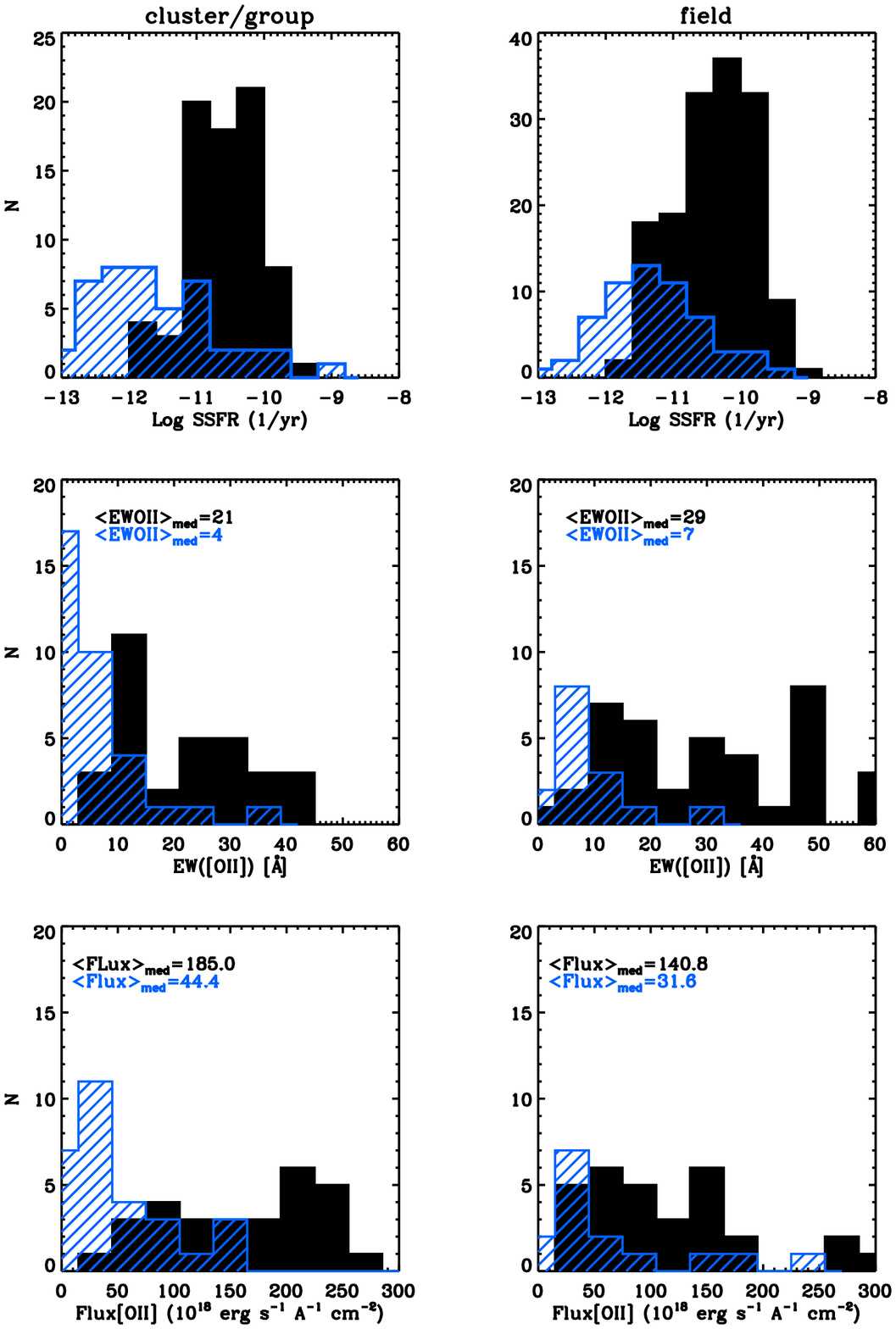}
\end{center} 
\caption{A comparison between the star formation of the kinematically-disturbed galaxies (shaded blue histograms) and the undisturbed ones (solid black histograms). Cluster galaxies are shown in the left hand panels and field galaxies in the right. The top row shows the specific star formation rates, the middle the equivalent width of the [OII] emission, and the bottom row compares the flux in OII. Median values are shown inside the plots.}
 \label{sfplots}
\end{figure}

In Sections~\ref{subsec:TF} and \ref{subsec:groups}, we found that the TFR of ``good'' galaxies (i.e. galaxies with no sign of kinematical distortion) is not significantly affected by environment. To test the effect that environment may have on the  kinematically-disturbed galaxies, which cannot be placed on the TFR, we take a more direct route by comparing the specific star formation rates (SSFRs, see Section~\ref{subsec:SFRs}) of the kinematically-disturbed galaxies with the rest. 
We find that kinematically-disturbed galaxies show lower SSFRs than their non-disturbed counterparts in all environments. This is shown in the top row of Figure~\ref{sfplots}. 
The KS statistics yield a very small probability  that the two samples (kinematically disturbed and undisturbed) follow the same distribution ($P_{\rm KS}$ of the order of $10^{-14}$), which means that this distributions are certainly not the same. Our sample exhibits a lower SSFR  for the kinematically-disturbed galaxies, particularly in cluster environments.

In Section~\ref{subsec:badfractions}, we showed that there are more kinematically disturbed
galaxies in clusters and groups than in the field, and therefore our finding
is consistent with that of \citet{Poggianti2008}, who showed that cluster
galaxies have slightly lower average SSFR than field ones. The suppressed
SSFR for the kinematically-disturbed galaxies is also seen in the field, so it
is not exclusively a cluster phenomenon. However, since there are more
disturbed galaxies in clusters than in the field, the average SSFR of starforming
cluster galaxies is smaller than that of field ones, in agreement
with \citet{Poggianti2008} results.

Although the difference in the SSFR distributions of disturbed and
undisturbed galaxies is very clear, there is a potential caveat. If a galaxy
has a low SSFR it will have a low [OII] emission line equivalent width (EW). This will make fitting the rotation curve more difficult, lowering the quality of the fits, and increasing the probability that the galaxy is classified as kinematically disturbed. In the middle and bottom panels of Figure~\ref{sfplots} we compare the EW and flux of the [OII] doublet for the ``good''
and ``bad'' galaxies in clusters and in the field separately. We find that ``bad'' or kinematically-disturbed galaxies have lower EW[OII] and lower [OII] flux in all environments. The problem arises when trying to decide
which is the cause and which the effect. The perturbed gas kinematics could be related to a process that also suppresses the SFR, providing a real physical link between both observations. However, it could also be that low SSFR galaxies have lower [OII] fluxes and EWs, making their rotation curves more difficult to fit well, and thus the apparent link is purely observational and not physical. Using only the information presented in this paper so far it is very difficult to know which one of these possibilities is the true one. However the additional independent evidence indicating that star formation is suppressed in cluster starforming
galaxies \citep{Poggianti2008,Vulcani2010,Finn2010} suggests that the observed connection between disturbed
kinematics and suppressed SSFR is a physical one. The results of Section~\ref{subsec:sizes} will also support this conclusion.

\subsection{Concentration of the emission}
\label{subsec:sizes}

To examine the location of the star formation within the disks of our emission-line galaxies, and its dependence on environment, we compared the size of the stellar disk, as traced by the photometric scale-length ($r_{\rm d,phot}$), with the size of the gas disk, i.e. the scale-length of the emission lines in the spectra ($r_{\rm d,emission}$). The emission-line scale-lengths were an output from the fits performed with ELFIT2PY, as described in Section~\ref{subsec:RC}. Photometric scale-lengths were derived by fitting a a 2-component 2D model that accounted for a bulge with a de Vaucouleurs profile and an exponential disk component, convolved with the PSF of the images. This was done using the GIM2D software \citep[see][for a detailed description of the method used]{Simard2002,Simard2009}. The values of $r_{\rm d,phot}$ used here were computed from the HST F814W  images, because of the higher quality of the data. We note however that if we used the scale-lengths measured from $I$-band VLT photometry, the results presented here would not change. We note that many dynamically hot systems have simple exponential profiles, hence the presence of a ``disk'' component does not necessarily imply the presence of an actual disk. For this reason, in this section we only considered galaxies that have been visually classified as disks (S0s and spirals only).

\begin{figure}
\begin{center}
  \includegraphics[width=0.5\textwidth]{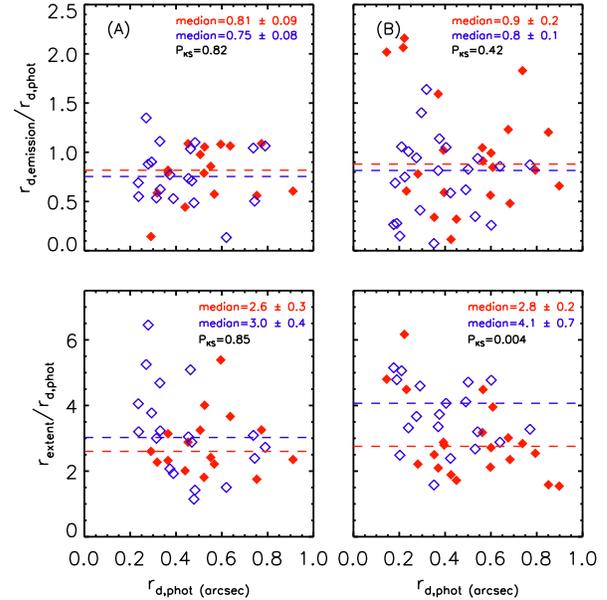}
\end{center} 
\caption{A comparison of the scale-lengths measured in the emission-lines ($r_{\rm d,emission}$, top panel) and the  emission-line extent($r_{\rm extent}$, bottom panel) versus those obtained from the photometry ($r_{\rm d,phot}$) in different environments. Only kinematically ``good'' galaxies  with disk morphology (S0s and spirals) in the matched samples A (left) and B (right) were taken into account. Cluster/group galaxies are plotted in filled red symbols, whilst field galaxies correspond to the open blue diamonds. 
The red and blue dashed lines show respectively the median deviation from a flat distribution, for the cluster/group and  field galaxies, respectively. The quoted uncertainties represent 1$\sigma$ errors in the median values (i.e. $1.253\times\sqrt{\rm rms}/\sqrt{\rm number~ of~ points}$, slightly larger than the error on the mean, but more robust to outliers). These plots show that whilst there is no difference in the location of the star formation within the stellar disks of cluster/group and field galaxies (top), there seems to be a truncation of the gas disks in cluster/group galaxies with respect to the field (bottom).}
 \label{rdplots}
\end{figure}

In Figure~\ref{rdplots}, we compare both scale-lengths. The top panels show the ratio $r_{\rm d,emission}/r_{\rm d,phot}$ plotted against $r_{\rm d,phot}$ in the mid- and high-redshift samples (A and B, respectively) for cluster/group and field galaxies in different symbols.  
The median values of this scale length ratio are the same (within the errors) for cluster/group and field galaxies in both samples. This suggests that the environment is not significantly affecting the gas concentration in emission-line galaxies that show no evidence of kinematical distortions. 

In contrast with this result, \citet{Bamford07} found that the emission (and thus the star formation)
of cluster spirals seems to be more concentrated that that of field ones.
Since these authors did not separate kinematically undisturbed and 
disturbed galaxies we repeated the test using all our fits, ``good'' and
``bad''. In this case we did find some weak evidence suggesting a more
concentrated star formation in cluster galaxies than in field ones, but the
large scatter introduced by the unreliable values of $r_{\rm d,emission}$ derived
from the ``bad'' fits prevented us from reaching any definitive conclusion.

When fitting the emission lines with ELFIT2PY (Section~\ref{subsec:RC}), the extent of the line, $r_{\rm extent}$, is also computed. This quantity is defined as the distance from the continuum centre to where the line could no longer be
reliably detected above the noise. Although $r_{\rm extent}$ depends on properties of the data (e.g. seeing, pixel size) and is thus not suitable for comparison with other studies, it is useful for the internal comparison of our own dataset. 
We use this quantity to investigate whether the extent of the gas disk is affected by cluster environment. 

The bottom row of Figure~\ref{rdplots} shows how the extent of the emission compares to the size of the stellar disk in a similar manner as Figure~\ref{rdplots}. Despite the scatter, the plots exhibit a $\sim1-2\sigma$ difference between field and cluster/group galaxies. This is more evident in the higher redshift sample (B). Cluster/group galaxies show smaller emission extents than field galaxies, implying that the cluster environment effectively truncates the gas disks. This is consistent with the results of \citet*{Koopmann2004}, who,  found that $\sim50$\% of spiral galaxies in the Virgo cluster have their H$\alpha$ disks truncated, whereas field galaxies do not show such evidence as frequently. Additionally, they find that most of the galaxies that exhibit truncated gas disks have relatively undisturbed stellar disks. From their results, they conclude that the reduced SFRs of Virgo spiral galaxies must be mainly caused by ICM gas stripping, which is also the scenario that our results favour.

\begin{figure}
\begin{center}
  \includegraphics[width=0.5\textwidth]{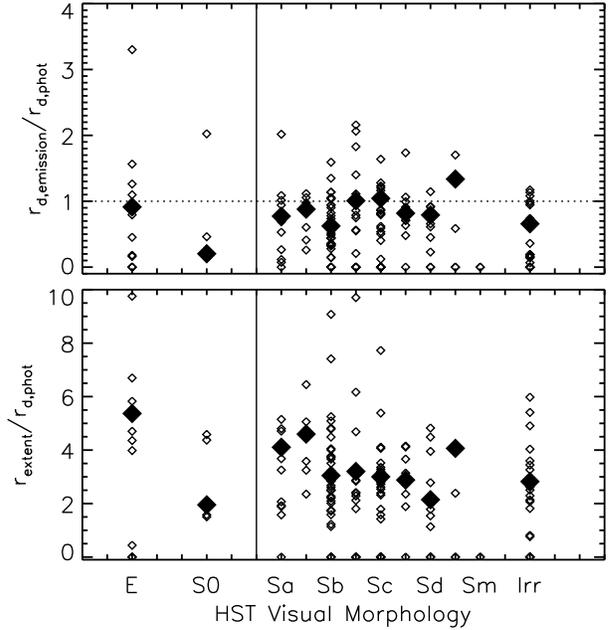}
\end{center} 
\caption{A comparison of $r_{\rm d,emission}/r_{\rm d,phot}$ (top) and $r_{\rm extent}/r_{\rm d,phot}$  with the different morphologies, for all the emission-line sample with HST observations and ``good'' emission-line fits. The horizontal dotted line in the top panel just guides the eye to where $r_{\rm d,emission}=r_{\rm d,phot}$ and the vertical solid line (both panels) divides early- from late-type galaxies. The larger solid symbols highlight the median values for each morphology type.}
 \label{rd_mor}
\end{figure}

In the top panel of Figure~\ref{rd_mor}, we plot the ratio $r_{\rm d,emission}/r_{\rm d,phot}$ as a function of morphology, for all the emission-line galaxies. We find that $r_{\rm d,emission}/r_{\rm d,phot}$  is roughly constant (with some scatter) throughout all the morphology types. The bottom panel shows $r_{\rm extent}/r_{\rm d,phot}$ for the different morphology types. A small decrease in  $r_{\rm extent}/r_{\rm d,phot}$ is observed towards later morphological types.  
If spiral galaxies transform into S0s in clusters one important issue is how to build the S0 bulges, since the average bulge-to-disk ratio of S0s is larger than that of spirals \citep{ChristleinZabludoff2004}. If the star formation is more concentrated in cluster spirals than in field ones, this will help to increase the bulge luminosity. This is consistent with 
%This is  consistent with 
the results shown in Figure~\ref{bad_frac_morph} and a scenario in which star forming spiral galaxies are transformed into passive S0s via stripping of their gas. 

%%%%%%%%%%%%%%%%%%%%%%%%%%%%%%%%%%%%%%%%%%%%%%%%%%%%%%%%%%%%%%%%%%%%%%
\section{Discussion}
\label{sec:discusion}
We have presented a detailed analysis of the effects of the environment on the gas and stars of distant galaxies. We have studied the gas kinematics, stellar morphology, Tully-Fisher relation, star formation, and concentration of the emission of galaxies in various environments, which has provided us with important clues about the physical mechanisms transforming galaxies. We summarize and discuss our results in the following.

From the full EDisCS galaxy sample at $z \lesssim 1$, we selected all galaxies with measurable emission in their spectra,  inclinations $>30^{\circ}$ (to avoid face-on galaxies), and slit misalignment (with respect to the major axis of the galaxy)  $<30^{\circ}$. We then modelled the 2D emission lines and fitted a rotation curve to obtain rotational velocities. All the fits were individually inspected in a quality check procedure that separated our galaxy sample into two categories. The first one contains galaxies for which their emission lines yielded acceptable fits (``good'' sample). The second one consists of galaxies for which no emission line could be fitted satisfactorily, and thus no reliable  rotational velocity could be derived (``bad'' sample). We then computed  $V{\rm rot}$ for each galaxy from the ``good'' emission-line fits. 

Galaxy morphology was not  taken into account in the sample selection. To investigate the morphology distribution of our emission-line sample, we studied the morphology distribution of the sub-sample of galaxies that have HST data (61\% of our sample). We found that while most of the emission-line galaxies in our sample are spirals, $\sim15$\% were classified as early-type galaxies (E or S0). Notably, the highest quality rotation-curve fits were obtained in the spiral sample, while the early-type galaxy group contained a significant fraction of ``bad'' galaxies. We nevertheless discovered 12 ellipticals and 5 S0s with clearly extended rotation curves in their emission. These interesting galaxies are being treated in a forthcoming paper (Jaff\'e et al.~in preparation).

We have shown that the galaxies with ``bad'' rotation curve fits represent a population of kinematically-disturbed galaxies. 
The fraction of kinematically-disturbed galaxies ($f_{\rm K}$) decreases significantly with morphological type (towards later types).  Within the spiral sample, there is a difference of a factor of $\sim 3$ between Sa and Scd galaxies, and this difference is even higher if we include S0 galaxies. In the context of spiral-to-S0 transformation, this implies that galaxies already having S0 morphology have been subject of stronger gas disturbance.

By studying the fraction of kinematically-disturbed galaxies over the same $M_{B}$  and redshift range in cluster, group, and field environments, we have found that $f_{\rm K}$ is clearly higher in cluster/group environments than in the field (for $M_{B}<-20.5$). The presence of kinematically-disturbed galaxies in clusters was first found by \citet{Rubin1999} in the Virgo cluster, and has been confirmed by other similar studies at higher redshift \citep[e.g.][]{Moran2007b, metevier06}. The difference in the kinematics between cluster and field galaxies we find for EDisCS emission-line galaxies agrees with theese previous results.

While the fraction of kinematically-disturbed galaxies in the field is roughly constant ($\cong 25$\%) throughout the $M_B$ range, in clusters $f_{\rm K}$ is not only higher, but  increases with luminosity. In other words, the most luminous (massive) galaxies exhibit more signs of gas disturbance. We interpret this trend as evidence that many of the fainter (less massive) galaxies have been completely stripped of their gas. This causes them to have no (or very little) emission in their spectra. For this reason, these galaxies are not selected in our emission-line galaxy sample. 
Moreover, we propose that if we were able to detect emission in these galaxies, the fraction of cluster galaxies with disturbed gas kinematics should be significantly higher than in the field at all luminosities, with a much smaller luminosity dependence. We have considered, but disfavour, two alternative explanations for the observed behaviour. First, the most luminous galaxies could be those that were accreted more recently and therefore their observed properties will reflect the recent influence of the cluster environment. This could be the result of the hierarchical cluster assembly, where more massive systems are accreted later (De Lucia et al.
in preparation). This interpretation is unlikely because we also find that the fraction of kinematically-disturbed galaxies decreases with projected distance from the cluster centre (see below). Second, it could be that fK grows with luminosity because brighter emission line galaxies may reside at the centre of the clusters, where we find higher incidence of kinematically-disturbed galaxies. We discard this possibility because we find no correlation between the luminosity of our cluster emission-line galaxies and their distance to the cluster centre.

To ascertain which physical mechanisms are affecting the gas kinematics, we studied how $f_{\rm K}$ varies with different proxies for environment. We found that, although $f_{\rm K}$ increases with cluster velocity dispersion (by a factor of $\sim2$) and decreases with distance from the cluster centre (by the same factor),  it remains constant with projected galaxy density. Although our results suffer from considerable uncertainty, they are self-consistent, and suggest that the physical mechanism acting on cluster galaxies is probably related to the ICM or the cluster potential itself and not to galaxy interactions.

We also tested whether there is any correlation between the degree of kinematical disturbance in the galaxies' gas and the amount of
disturbance in their morphologies.
We did this by fitting a smooth single-Sersic index model to each galaxy (with available HST data) and subtracted it from the original HST image. The corresponding residual images thus highlighted morphological distortions. By inspecting them carefully, we found that  $\sim50$\% of the galaxies show  signs of asymmetry that we have interpreted as the possible result of a recent interaction (or merger event in the most dramatic cases). 
We did not find a clear direct link between the kinematic disturbance in the galaxies' gas and their morphological disturbance, indicating that the physical mechanisms and/or timescales involved are different.”

We then searched for environmental effects on the galaxies' scaling relations, by studying 
the Tully-Fisher relation of cluster, group and field galaxies. We only considered kinematically non-disturbed (``good'') galaxies within matched samples (in $M_B$ and z). We found that there is no difference between the distribution of cluster, group and field galaxies in the Tully-Fisher diagram up to $z<1$. The distributions are strikingly similar. 
This result agrees with \citet{Nakamura2006} and \citet{ziegler03} but contradicts the findings of \citet{Bamford05}, who found a brighter TFR for cluster galaxies. Because our sample is larger and more homogeneous than the one published by these authors, and our quality control more robust we are confident on the reliability of our findings. Taken at face value this result suggests that the cluster environment does not induce a strong enhancement on the star-formation activity of spiral galaxies entering it.

In an attempt to reduce the scatter about the TFR, we have performed the above-mentioned analysis with only morphologically confirmed spirals. This reduced the number of galaxies significantly (by half) since we do not have HST observations for all the emission-line sample. Nevertheless, we obtained a tighter TFR  \citep[as expected, e.g.][]{Kannappan2002} and were able to make comparisons between the different environments. Our results show that, for the spiral sample, the cluster/group TFR  again does not differ significantly from the field relation. 
No statistically-significant difference is found either when comparing the TFRs of galaxies in the field and in clusters with $\sigma_{cl}>400$km/s (i.e., when excluding group galaxies). 

To further confirm that the TFR is not significantly affected by environment, we studied the TFR residuals as a function of cluster velocity dispersion, projected distance from the cluster centre and projected galaxy density, and found no evidence for a correlation between environment and TFR residuals.

At face value, the fact that we find no significant environmental effects on the TFR seems to suggest that there is no strong enhancement or suppression of the star formation activity in cluster star forming spiral galaxies. 
However this cannot be the whole story, since the TFR analysis can only be properly done for galaxies with reasonably regular rotation curves (and thus galaxies without strong distortions in their gas structure and kinematics). 
If the main environmental effects on spirals manifest themselves as disturbances in the galaxies' gas, the kinematically-disturbed galaxies are a key component of the whole picture. Because these galaxies cannot be reliably placed on the TFR we need to use other tests to assess the effect of the environment on their star formation activity.
Using the [OII] emission line as an estimator of the galaxies' current star formation we find that kinematically-disturbed galaxies exhibit lower specific star formation rates (SSFR, i.e., star formation rate per unit stellar mass) in all environments. Although some observational biases may be at play, using independent evidence from previous EDisCS studies we argue that this effect is probably real. If so, this suggests that there may be a physical connection between the disturbance in the galaxies' gas and their reduction in star-formation activity.

Further support to this interpretation comes from our study of the spatial distribution of the line emission, taken as a tracer of star formation. The concentration of the star formation, parameterised as the ratio of the exponential scale length of the line emission divided by the exponential scale length of the stellar disk, seems to be unaffected by the environment for the galaxies with undisturbed gas. However, although the exponential scale lengths of the line emission do not seem to be affected, the actual extent of the emission appears to be. The radial extent of the galaxies' emission (in units of their stellar disk scale length) is smaller in cluster environments than in the field. In other words, the star formation seems to be more concentrated (or truncated) in cluster galaxies. This means that the cluster environment not only reduces the galaxies star formation activity but also makes what star formation remains more concentrated. This has been independently observed in clusters at lower redshifts \citep{Wolf2009}.

\section{Conclusions}
\label{sec:conclusions}

We have studied the properties of the gas and the stars in a sample of 422 emission-line galaxies from the ESO Distant Cluster Survey in different environments at $0.3<z<0.9$. Our principal aim is to try to understand the main physical mechanisms acting on galaxies when they fall into clusters, Our main findings are:

\textbf{(i)} The fraction of galaxies with kinematically-disturbed gas disks is higher in galaxy clusters than in the field. While this fraction does not change with luminosity in the field, in clusters it increases significantly with increasing luminosity. We can explain this trend as the consequence of gas being more easily removed from lower mass (fainter) galaxies, taking them out from the emission-line galaxy sample.

\textbf{(ii)} The fraction of kinematically-disturbed galaxies increases with cluster velocity dispersion and decreases with projected distance from the cluster centre, which is indicative of strong environmental effects on the galaxies' gas. However, we found no correlation between the fraction of kinematically-disturbed galaxies and the projected galaxy density. We interpret this as a strong indication that what is causing disturbances in the galaxies gas is likely related to the ICM and not due to galaxy-galaxy interactions.

\textbf{(iii)} The fraction of galaxies with disturbed optical morphologies in our emission-line sample is luminosity independent and similar in clusters, groups, and the field, Indeed, there is little correlation between the presence of kinematically-disturbed gas and morphological distortions. These results, combined with (i) and (ii) above, suggest that environmental effects are mild enough to ensure that, whilst they do not disturb the stellar disks, they do strongly affect the gas in cluster galaxies.

\textbf{(iv)} No environmental effects on the Tully-Fisher relation are found for the emission-line galaxy sample nor for the morphologically-classified spirals.

\textbf{(v)} Result (iv) is inevitably limited to the galaxies with undisturbed kinematics.  Since reliable rotation velocities cannot be determined for kinematically-disturbed galaxies, these cannot be placed of the Tully-Fisher relation. For this reason we explored the possibility that signatures of enhanced or suppressed star formation could be present in the kinematically-disturbed galaxies. Indeed, we find that kinematically-disturbed galaxies have lower specific star formation rates.

\textbf{(vi)} Cluster galaxies display truncated star-forming disks relative to similarly-selected field galaxies.

\textbf{(vii)} There are several galaxies that have been morphologically classified as E/S0, that exhibit extended gas disks. These galaxies will be discussed in a forthcoming paper.

Previous studies have shown that, statistically, spiral galaxies probably transform into S0s in cluster environments \citep[e.g.][and references therein]{Desai2007}. This fact, together with the results presented in this paper, lead to the following conclusions: if infalling spirals are the progenitors of cluster S0s, the physical mechanism responsible for this transformation is such that it efficiently disturbs the galaxies' star-forming gas and reduces their star-formation activity, but leaves their stellar disks largely undisturbed. Moreover, the star-forming gas is either removed more efficiently from the outskirts of the galaxies, or it is driven towards the centre (or both). In any case, this makes any remaining star formation more centrally concentrated, helping to build the bulges of S0s. We conclude that the physical mechanism responsible for the spiral-to-S0 transformation in clusters is related to the intra-cluster medium, with galaxy-galaxy interactions and mergers playing only a limited role. Of course, this does not imply that S0s in lower-density environments cannot form via different mechanism(s).

%%%%%%%%%%%%%%%%%%%%%%%%%%%%%%%%%%%%%%%%%%%%%%%%%%%%%%%%%%%%%%%%

\section*{Acknowledgments}
Based on observations collected at the European Southern Observatory, Chile, as part of programme 166.A-0162. This work utilised the High Performance Computing facility at the University of Nottingham. YLJ is greatly indebted to the European Southern Observatory for the Studentship to work on this project, the astrophysics group of Exeter University for their hospitality, and the Royal Astronomical Society for their support. YLJ also thanks Bodo Ziegler for useful discussions.  
The Dark Cosmology Centre is funded by the Danish National Research Foundation. GDL acknowledges financial support from the European Research Council under the European Community's Seventh Framework Programme (FP7/2007-2013)/ERC grant agreement n. 202781.

%\bibliography{/home/yara/Desktop/thesis/style/refs}{}
%\bibliographystyle{mn2e}

\newpage

\appendix

%\newpage

\onecolumn 
\section[]{The data table}

Table~\ref{example_table} shows a shortened version (10 rows) of the full data table, available in the online version of this paper. The table contains our measurements of rotation velocity, kinematical disturbance,  and emission disk scalelengths, output from our 2D emission-line fitting procedure, as well as the morphological disturbances found from the single-Sersic fits to the HST data. We also included other characteristics  of the data for completeness.  
The table columns are: \\

%\begin{enumerate}
 (1)  Name of galaxy in the EDisCS catalogue. \\

 (2)  Galaxy environment: ``f'' stands for for field, ``c'' for cluster ($\sigma_{cl}\gtrsim400$km/s) and ``g'' for group ($\sigma_{cl}\lesssim400$km/s)\\

 (3)  Redshift.\\

 (4)  $B$-band magnitude corrected for internal extinction.\\

 (5)  Logarithm of the rotation velocity (derived from ELFIT2PY), and associated confidence error.\\

 (6) Inclination used (from HST photometry if available, otherwise computed from $I$-band VLT images).\\

 (7) Flag for kinematical disturbance (``good'' or ``bad'' for undisturbed and disturbed, respectively), as judged from the emission lines in the 2D spectra.\\

 (8)  Hubble T morphology type, obtained by visual inspection of the HST images. The numbers correspond to the following types: star=-7, X=-6, E=-5, S0=-2, Sa=1, Sb=3, Sbc=4, Sc=5, Scd=6, Sd=7, Sdm=8, Sm=9, Im=10, Irr=11, ?=66,  and ``-'' is placed whenever there is no HST data available.\\

 (9)  Flag for morphological disturbances (``good'' or ``bad'' for undisturbed and disturbed, respectively) as detected from the single-sersic fits made to the HST images. We note that these flags must be interpreted with care as they do not necessarily represent mayor morphological disturbances (cf. Section 5.3).\\

 (10)  The emission-line (exponential) disk scalelenght.\\

 (11)  Extent of the line as measured by ELFIT2PY (only usable within our data since it depends on e.g. seeing).\\

 (12)  The photometric disk scalelengths, obtained from HST data, plus their uncertainties.\\

 (13)  The photometric disk scalelengths, obtained from  VLT data, plus their uncertainties.\\

%\end{enumerate}
%
We note that the values of $\log V_{\rm rot}$, $r_{\rm d,emission}$, and  $r_{\rm extent}$ are not listed for kinematically disturbed galaxies (instead a ``--'' is placed), as these values are not physically correct and can thus be misleading.

\begin{landscape}
\begin{table}
\begin{center}
\caption{\label{example_table} 10 Example rows of the table containing all the measured quantities to the EDisCS emission-line sample (the full table can be found in the electronic version of the paper).  The columns are: (1) name of galaxy in the catalogue, (2) environment (``f'' for field, ``c'' for cluster and ``g'' for group), (3) redshift, (4) $B$-band magnitude corrected for internal extinction, (5) logarithm of the rotation velocity (from ELFIT2PY) and associated confidence error, (6) inclination used (from HST photometry if available, otherwise computed from $I$-band VLT images), (7) flag for kinematically disturbed (``bad'') or undisturbed (``good'') galaxies as judged by their emission-line fits, (8) Hubble T morphology type, obtained by visual inspection of the HST images (star=-7, X=-6, E=-5, S0=-2, Sa=1, Sb=3, Sbc=4, Sc=5, Scd=6, Sd=7, Sdm=8, Sm=9, Im=10, Irr=11, ?=66,  and ``-'' is placed whenever there is no HST data available), (9) flag for morphological disturbances (``good'' or ``bad'') as detected from the single-sersic fits made to the HST images,  (10) the emission-line (exponential) disk scalelenght, (10) extent of the line as measured by ELFIT2PY (only usable within our data since it depends on e.g. seeing), and (12 and 13) the photometric disk scalelengths (for HST and VLT data), plus their uncertainties.}
\begin{tabular}{ccccccccccccc}
\hline\\[-2mm]
Object ID &  envi-. &   $z$	&  $M_{B}$	&  $\log V_{\rm rot}$	 &  $inc$ &   kinem.	&   Hubble T &  morph. &  $r_{\rm d,emission}$ &  $r_{\rm extent}$   &  $r_{\rm d,phot}^{\rm HST}$ &  $r_{\rm d,phot}^{\rm VLT}$ \\ %\hline}

[EDCSNJ*] &  ronment &   	&  (mag)	&  (km/s)	&  ($^{\circ}$)  &  	 dist.	&  morph. &  dist. &  ($^{\prime\prime}$) &  ($^{\prime\prime}$) &  ($^{\prime\prime}$)&  ($^{\prime\prime}$)	\\

(1) & (2) & (3) & (4) & (5) & (6) & (7) & (8) & (9) & (10) & (11) & (12) & (13)\\
\hline
1119226-1128488   & f     & 0.5269        & $-$21.81      &		--		  &           39  & bad      & --    	&--   	&--      &--   & --    			&0.41$^{+0.09}_{-0.09}$ \\
1119235-1130144   & f     & 0.6777        & $-$20.85      &1.39$^{+0.234}_{-0.334}$       &           43  & good     & --    	&--   	&0.939$^{+0.339}_{-0.364}$      &1.14   & --    			&0.38$^{+0.02}_{-0.02}$ \\
1119243-1131232   & f     & 0.2125        & $-$20.52      &2.17$^{+0.019}_{-0.020}$       &           59  & good     & --    	&--   	 &0.607$^{+0.097}_{-0.097}$      &2.67   & --    			&1.05$^{+0.01}_{-0.01}$ \\
1138034-1132394   & f     & 0.6199        & $-$19.87      &1.16$^{+0.468}_{-100}$     	  &           58  & good     &  3  	&bad   	 &0.085$^{+0.033}_{-0.027}$      &3.40   & 0.13$^{+0.01}_{-0.00}$        &2.06$^{+0.13}_{-0.24}$ \\
1138035-1132254   & c     & 0.4785        & $-$20.83      &2.22$^{+0.046}_{-0.052}$       &           66  & good     &  5    	&good  	 &0.474$^{+0.054}_{-0.054}$      &1.30   & 0.45$^{+0.00}_{-0.00}$        &0.54$^{+0.01}_{-0.01}$ \\
1138037-1137275   & f     & 0.7384        & $-$21.71      &1.62$^{+0.165}_{-0.194}$       &           82  & good     &  11     	&good  	 &0.207$^{+0.027}_{-0.028}$      &1.03   & 1.27$^{+0.17}_{-0.35}$        &0.54$^{+0.06}_{-0.07}$ \\
1138057-1131517   & f     & 0.3586        & $-$19.02      &1.76$^{+0.131}_{-0.261}$       &           43  & good     &  6      	&bad   	 &0.224$^{+0.034}_{-0.035}$      &1.30   & 0.31$^{+0.01}_{-0.01}$        &0.34$^{+0.01}_{-0.01}$ \\
1138064-1134252   & f     & 0.6192        & $-$20.30      &2.15$^{+0.028}_{-0.023}$       &           36  & good     &  3      	&bad   	 &0.428$^{+0.006}_{-0.007}$      &1.40   & 0.38$^{+0.00}_{-0.01}$        &0.43$^{+0.02}_{-0.01}$ \\
1138064-1134297   & f     & 0.5452        & $-$19.31      &1.41$^{+0.222}_{-0.546}$       &           46  & good     &  11      &bad   	 &0.245$^{+0.010}_{-0.012}$      &1.27   & 0.26$^{+0.01}_{-0.01}$        &0.31$^{+0.03}_{-0.04}$ \\
1138069-1136160   & c     & 0.4520        & $-$18.62      &		--		  &           51  & bad      &  $-2$   	&bad   	 &--      &--  & 0.25$^{+0.01}_{-0.01}$        &0.24$^{+0.02}_{-0.02}$ \\
\hline
\end{tabular}\\
\end{center}
\end{table}
\end{landscape}

\end{document}